\documentclass[11pt,a4paper]{article}
\pdfoutput=1
\usepackage{jinstpub}
\usepackage[english]{babel}
\usepackage{upgreek}
\usepackage{xspace}
\usepackage{graphicx}
\usepackage[capitalize]{cleveref}
\usepackage{enumitem}
\usepackage{subcaption}
\usepackage[modulo]{lineno}

\usepackage{subcaption}
\usepackage[modulo]{lineno}

\title{Calibration of the underground muon detector of the Pierre Auger Observatory}

\author{\includegraphics[height=30mm]{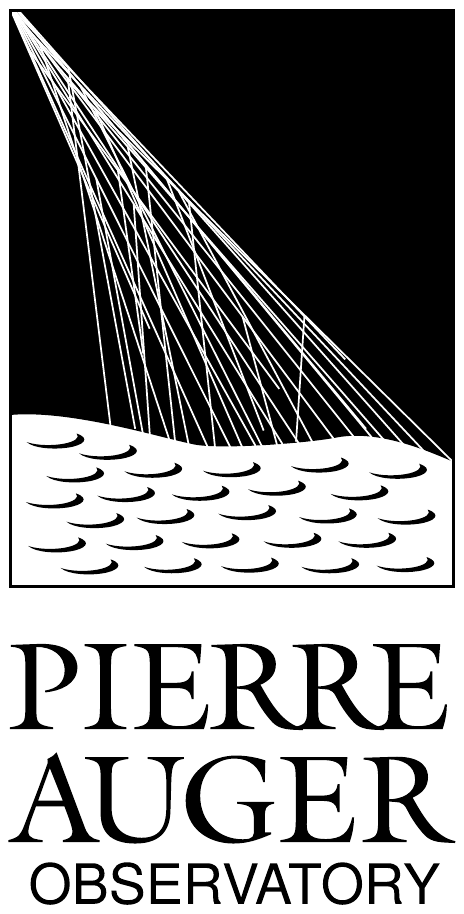}\\[3mm]The Pierre Auger Collaboration}
\affiliation{Av.\ San Mart\'{\i}n Norte 306, 5613 Malarg\"ue, Mendoza, Argentina}

\emailAdd{auger\_spokespersons@fnal.gov}

\abstract{To obtain direct measurements of the muon content of extensive air showers with energy above $10^{16.5}$\,eV, the Pierre Auger Observatory is currently being equipped with an underground muon detector (UMD), consisting of 219 10\,$\mathrm{m^2}$-modules, each segmented into 64 scintillators coupled to silicon photomultipliers (SiPMs). Direct access to the shower muon content allows for the study of both of the composition of primary cosmic rays and of high-energy hadronic interactions in the forward direction. As the muon density can vary between tens of muons per m$^2$ close to the intersection of the shower axis with the ground to much less than one per m$^2$ when far away, the necessary broad dynamic range is achieved by the simultaneous implementation of two acquisition modes in the read-out electronics: the binary mode, tuned to count single muons, and the ADC mode, suited to measure a high number of them. In this work, we present the end-to-end calibration of the muon detector modules: first, the SiPMs are calibrated by means of the binary channel, and then, the ADC channel is calibrated using atmospheric muons, detected in parallel to the shower data acquisition. The laboratory and field measurements performed to develop the implementation of the full calibration chain of both binary and ADC channels are presented and discussed. The calibration procedure is reliable to work with the high amount of channels in the UMD, which will be operated continuously, in changing environmental conditions, for several years.
}


\arxivnumber{2012.08016} 

\dedicated{\flushleft\textnormal{\textsc{%
Published in JINST as \href{https://dx.doi.org/10.1088/1748-0221/16/04/P04003}{DOI:10.1088/1748-0221/16/04/P04003}
\\
Report number: FERMILAB-PUB-20-678-AD-E-TD
}}}

\begin{document}
\maketitle
\flushbottom

\section{Introduction}
\label{sec::intro}

Many questions remain open concerning the origin of high- and ultra-high-energy cosmic rays. Due to their low flux, cosmic rays with energy above $10^{15}$\,eV can only be indirectly detected by measuring extensive air showers produced when they interact with molecules in the Earth's atmosphere. Therefore, the nature of each primary cosmic ray, or the composition of their global flux, is difficult to determine, hindering the full understanding of data. However, the total number of muons produced in air showers scales with the mass number of the primary particle. Direct measurements of these muons are thus optimal to understand the chemical composition of primary cosmic rays.

The aim of the Pierre Auger Observatory~\cite{PierreAugerObservatory}, located near the town of Malargüe in Argentina, is to measure cosmic rays with energies above $10^{17}$\,eV. A hybrid technique is employed to do so, in which a fluorescence detector (FD) and a surface detector (SD) are used in combination to detect extensive air showers. The FD is composed of 27 fluorescence telescopes distributed at four sites at the edge of the SD. The SD consists of an array of 1600 water-Cherenkov detectors (WCDs) with a 1500\,m spacing (SD-1500), covering a total area of 3000\,km$^2$, and a denser array of 71 WCDs over 28\,km$^2$ with a 750\,m spacing (SD-750). More recently, the SD energy threshold has been extended down to $10^{16.5}$\,eV thanks to the installation of an even denser array of about 2\,km$^2$ in which detectors are spaced by 433\,m (SD-433).

To provide a direct measurement of the muon content in air showers, an underground muon detector (UMD)~\cite{SDInfill,AugerPrime} is being deployed at the Auger site. The UMD will consist of an array of 219 scintillator modules co-located at 73 positions: 61 positions of the SD-750, forming a compact hexagon, and at each position of the SD-433. At a distance of at least 7\,m from the WCD, three 10\,m$^2$ scintillation modules are buried at a depth of 2.3\,m. The separation from the surface detector guarantees that even particles with zenith angles up to 45$^\circ$ hit the scintillators without passing through the WCD while probing the same shower density. The chosen depth, which corresponds to 540 g/cm$^2$ of overburden as determined by the local soil density, ensures that the electromagnetic component of extensive air showers is largely absorbed while vertical muons with energy >1\,GeV can reach the buried detectors. Each underground module is segmented into 64\,plastic-scintillation strips of 400\,cm$\times$1\,cm$\times$4\,cm with embedded wavelength-shifting optical fibers coupled to an array of 64 silicon photomultipliers (SiPMs)~\cite{AMIGAPrototype, AMIGASIPM}, Hamamatsu S13361-2050.

The acquisition of the UMD data is performed in slave-mode, this is, only when triggered by the associated SD stations. The shower geometry (arrival direction and intersection of the shower axis with the ground) and the energy are reconstructed with data from the WCDs alone by analyzing the signal as a function of their transverse distance to the shower axis~\cite{SDEventReconstruction}. The number of muons at the ground is, on the other hand, reconstructed with the UMD data separately. A global estimator of the shower muon density is obtained by reconstructing the muon lateral distribution~\cite{AMIGALDF} from all UMD stations and obtaining its value at 450\,m (the distance at which fluctuations of signals in individual showers measured with the SD-750 array are minimized with respect to an average lateral distribution). In a previous work, using data of the UMD engineering array, we confirmed that the number of muons in data is larger than those expected from Monte-Carlo simulations of showers between $10^{17.5}$\,eV and $10^{18}$\,eV, as reported by other experiments~\cite{MuonDeficit, MuonsWithAMIGA}, which shows that the models used in simulations do not describe accurately the high energy hadronic interactions in the forward direction. After the engineering-array phase, modifications in the design were implemented to optimize the detector performance based on the experience gained with the operation of prototypes. In particular, photomultiplier-tubes were replaced with SiPMs to improve the detection efficiency, and upgraded electronics with an additional acquisition mode to extend the dynamic range were implemented~\cite{UMDFrontEnd}.

A key element for the reconstruction of the lateral distribution function is the conversion of UMD signals in units of muons, which relies on the calibration described in this work. The density of muons at the ground depends on the energy, primary composition, zenith angle, and distance from the shower core: it can vary between tens of muons per m$^2$ close to the core to much less than one per m$^2$ when far away. Therefore, the UMD needs to work efficiently in a broad dynamic range and, to achieve this, two complementary working modes are implemented: one, dubbed {\it binary}, is tuned to count single muons, the other, named {\it ADC}, is devoted to measuring a high number of them. We describe in detail these two modes in section~\ref{sec:electronics}, where we also discuss how the signal acquired in both modes correlates.

The way to convert raw signals into a number of muons is different for the two acquisition modes. In the binary mode, the identification of particles crossing the detector relies on its segmentation, and the number of muons is determined by counting signals, segment-by-segment, with an amplitude higher than a given threshold. In contrast, the ADC mode relies on the determination of the charge of all segments simultaneously, converted afterward to a number of muons through the average charge of the signal produced by a single muon. The calibration methods for the two modes are explained in Sections~\ref{sec:SiPmcalibration} and ~\ref{sec:integrator}, respectively. Finally, we summarize and conclude in section~\ref{sec:conclusion}.

\section{Electronics features of the underground muon detector}
\label{sec:electronics}

The UMD electronics consist, basically, of: (i) an array of SiPM with its read-out system, (ii) a Field-Programmable Gate Array (FPGA) that houses the acquisition logic and a soft-core for data transmission and slow control, and (iii) an interface for the WCD trigger, monitoring data, and communications. Working as a slave-detector, the event acquisition in the UMD stations is governed by that of the SD, whose trigger chain has different levels~\cite{PierreAugerTrigger}, being the first (T1) and second (T2) ones formed locally at each WCD. The averaged rate of T1s and T2s is 100 and 20\,triggers\,per\,second, respectively. The buried scintillators are synchronized at the first-level trigger: when such a condition is met, the WCD sends a signal with a timestamp to all the underground detector modules associated with it. Upon the arrival of such a signal, the traces of both, binary and ADC, acquisition modes, as well as the timestamp of the trigger, are stored in an internal memory that can accommodate data up to 2048 triggers. Therefore, the rate of triggers determines that, at each position, the data are retrievable up to about 20\,s after its occurrence. Most of the T1s in the WCD are triggered by background muons and, as such, uncorrelated among the array. On the other hand, all WCDs passing the second-level triggers send their timestamp to a central data acquisition server to form a global shower-trigger that initiates the data acquisition from the array for permanent storage. When a shower trigger takes place, all first- and second-level triggers from the WCDs, along with the muon detector traces, are sent to the central data acquisition server.

A broad dynamic range is attained by the implementation of two acquisition modes, binary and ADC, which work simultaneously. The binary mode benefits from the detector segmentation: muons can be directly counted as pulses above a certain threshold. This mode is very robust since it neither relies on deconvoluting the total number of particles from a single integrated-signal, nor on the precise optical device gain or its fluctuations, and is almost completely independent of the
hitting position of the particle on the scintillator strip and the corresponding light attenuation through the fiber. Another advantage is that it does not require a thick scintillator to control Poissonian fluctuations in the number of photons per impinging muon. Nevertheless, the binary acquisition mode is limited by the segmentation itself: two muons arriving at the same strip simultaneously will be indistinguishable and, therefore, only counted once. This feature is known as {\it pile-up} effect. As far as strips with a simultaneous signal are less than the segmentation, the pile-up effect can be statistically treated and corrected~\cite{MuonsCounters}, but as soon as the 64 strips in a given module have signal at the same time, the correction is no longer feasible and the binary mode is saturated. If the pile-up effect might produce under-counting, knock-on electrons from the soil or inclined muons might produce signals in two neighboring strips resulting in an over-count of particles as well as an increased probability of saturating the detector. Therefore, the binary mode is limited in the number of muons that can be detected at the same time, which translates into a limit in the distance to the shower core that can be probed with it. Furthermore, since the fluctuations in the number of hit segments depend on the number of impinging particles, the resolution of the binary mode worsen when measuring closer to the shower core.

The ADC mode extends the saturation point of the detector: it is well-suited to measure a higher number of particles using the integrated signals of all 64 strips as a single measurement of the total charge in the module. Contrary to the binary acquisition mode for which fluctuations grow with the number of impinging particles, the fluctuations in the ADC mode decrease with the number of detected muons. However, since it relies on estimating the number of particles from an integrated signal in terms of an average charge, the signal fluctuations are propagated to uncertainties in the estimation of the number of muons. The resolution of the ADC channel is 60\% for single-muon signals and decreases inversely to the square root of the number of particles; it reaches 10\% at a few tens of muons where it matches the resolution of the binary mode. At higher muon numbers, the ADC mode has significantly better resolution than the binary mode, allowing for measurements of particle densities closer to the core with improved precision. 
In the binary acquisition mode, the 64 SiPM signals are handled independently through a pre-amplifier, fast-shaper, and a discriminator, built within each channel of two 32-channel Application-Specific Integrated Circuits (ASICs). The discriminator signal is sampled at 320\,MHz (3.125\,ns sample time) by the FPGA into 64 traces of 2048 bits. In each trace, a ``1''-bit results if the fast-shaper output is above a fixed discriminator threshold, and a ``0''-bit otherwise. As explained in section~\ref{sec:SiPmcalibration}, this threshold is defined to significantly reject the SiPM noise while preserving a high detection efficiency.

\begin{figure}[t]
	\centering
	\begin{subfigure}{0.45\textwidth}
		\centering
        \includegraphics[trim={0.0cm 0.0cm 39.5cm 0.0cm},clip,width=1.0\linewidth]{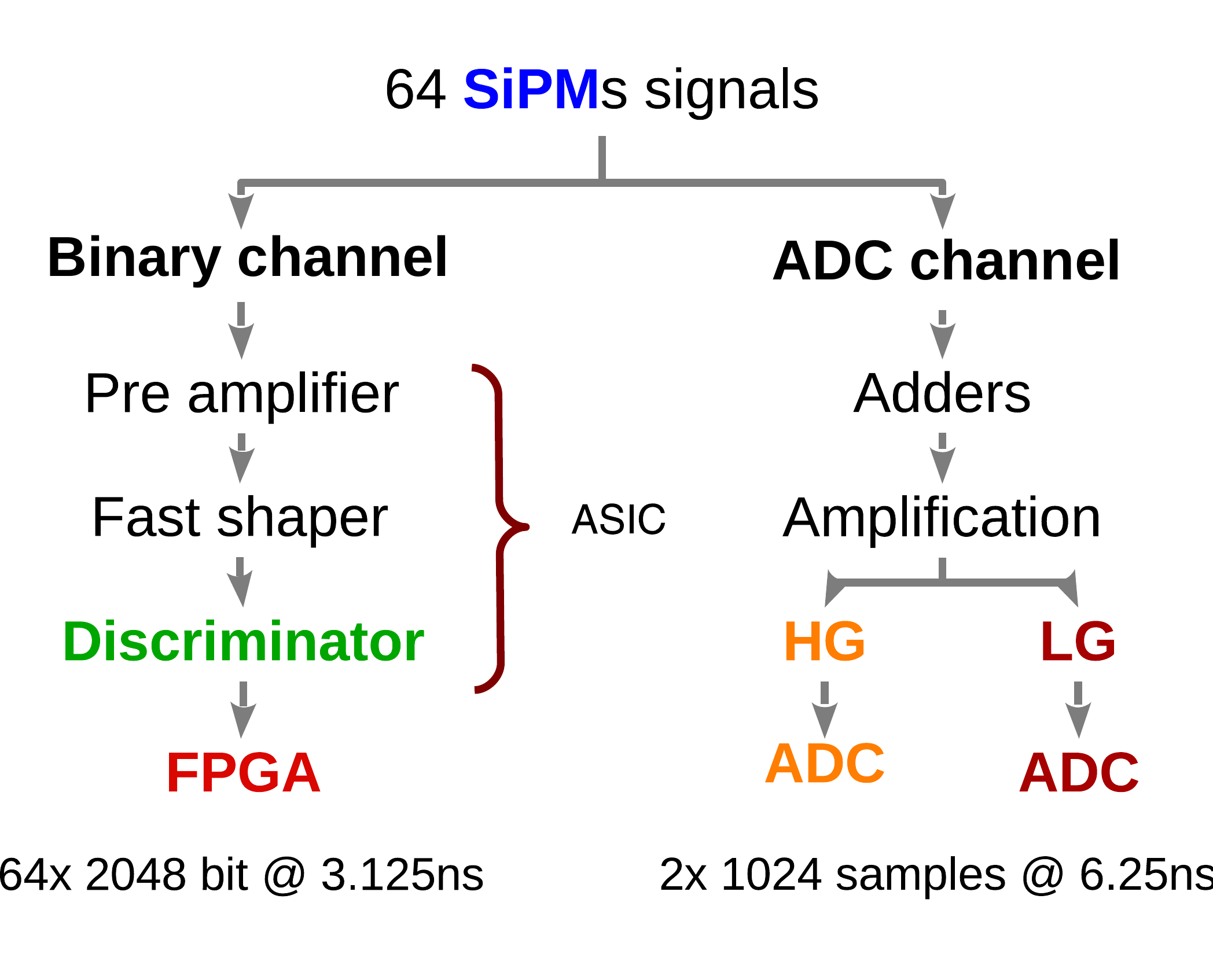}
	\end{subfigure}\\
	\centering
		\begin{subfigure}{0.7\textwidth}
		\centering
		\includegraphics[trim={0.0cm 0.0cm 0cm 0.0cm},clip,width=1.0\linewidth]{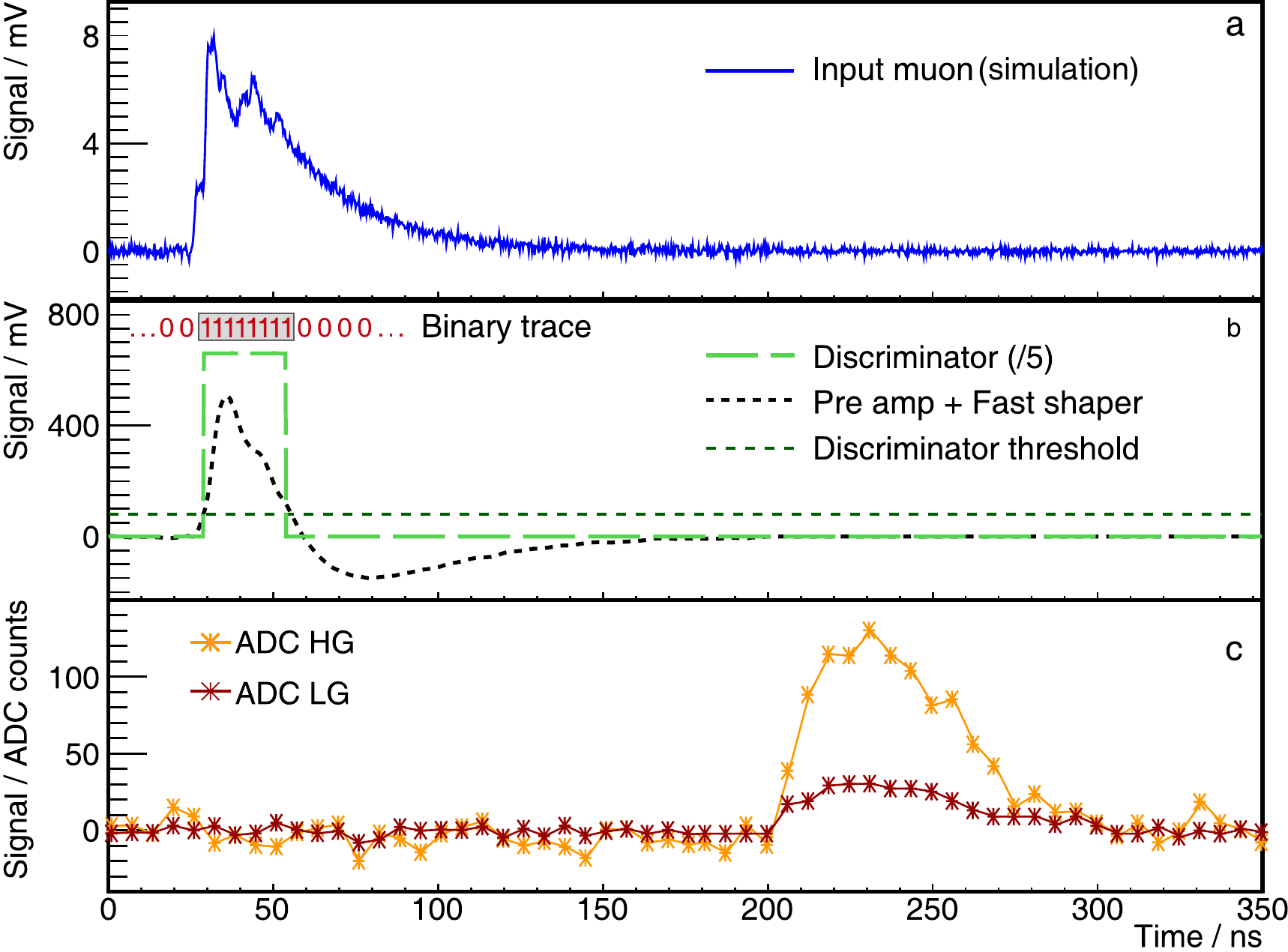}
	\end{subfigure}
    \caption{(Top) Schematics of the UMD electronics. (Bottom-a) Simulated SiPM response to a single-muon signal. (Bottom-b) Simulation of the binary mode response to the simulated muon. The discriminator pulse is re-scaled for illustration. (Bottom-c) Simulation of the ADC mode processing. The two ADC outputs (low- and high-gains) are shown. 
    \label{fig:electronicsSchematic}}
\end{figure}

In the ADC acquisition mode the 64\,SiPM signals are added up analogically and the result is amplified with low- and high-gain amplifiers with a ratio of about 4 in the amplification factor. The low-gain channel extends the dynamic range of the UMD to measure a higher number of muons when the binary mode is saturated. The high-gain channel, initially conceived to work in an intermediate range of muon numbers, resulted to perform similarly to the binary channel. The amplifier outputs are sampled at 160\,MHz (6.25\,ns sample time) with two 14-bit Analog-to-Digital Converters (ADCs) resulting in two waveforms of 1024\,samples. A schema of the read-out electronics is shown in the upper panel of Fig.~\ref{fig:electronicsSchematic}. In the lower panel, we show, from top to bottom, how the SiPM (a), the binary channel (b), and the ADC channel (c) respond to a simulated single muon.

As the signal processing in binary and ADC modes is performed with different sets of electronics, differences in the response to the same physical event are expected. The different architectures introduce, in particular, different read-out delays, so that the start time of the signals in binary and ADC channels do not coincide: as one can see in the lower panel of Fig.~\ref{fig:electronicsSchematic}, the start-time of the ADC trace is delayed with respect to the binary one. Nevertheless, as we shall shortly see in the following, this time shift is very-well defined, stable in time, and uniform across the different electronics.

\begin{figure}[t]
	\centering
		\begin{subfigure}{0.8\textwidth}
		\centering
		\includegraphics[trim={0.0cm 0.0cm 0cm 0.0cm},clip,width=1.0\linewidth]{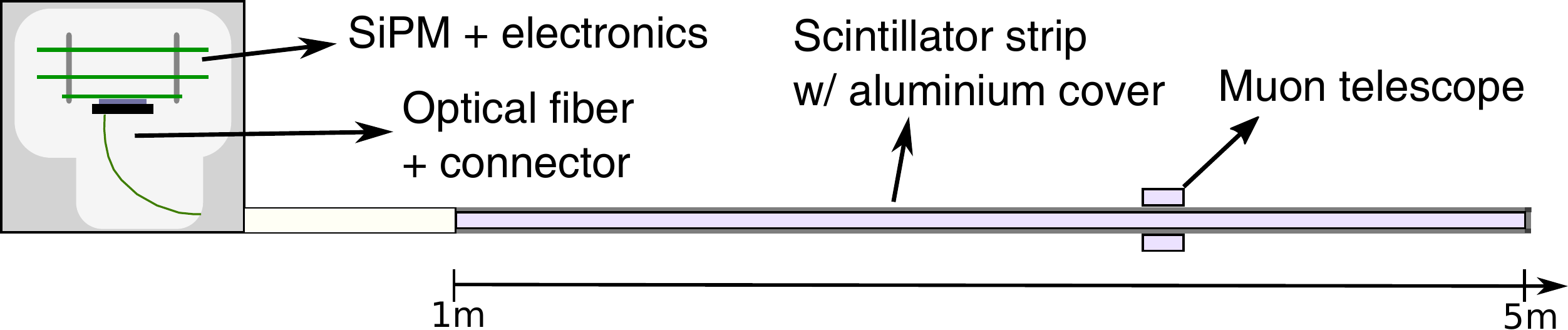}
	\end{subfigure}\\
	\vspace{0.5cm}
	\centering
		\begin{subfigure}{1.00\textwidth}
		\centering
		\includegraphics[trim={0.0cm 0.0cm 0cm 0.0cm},clip,width=1.0\linewidth]{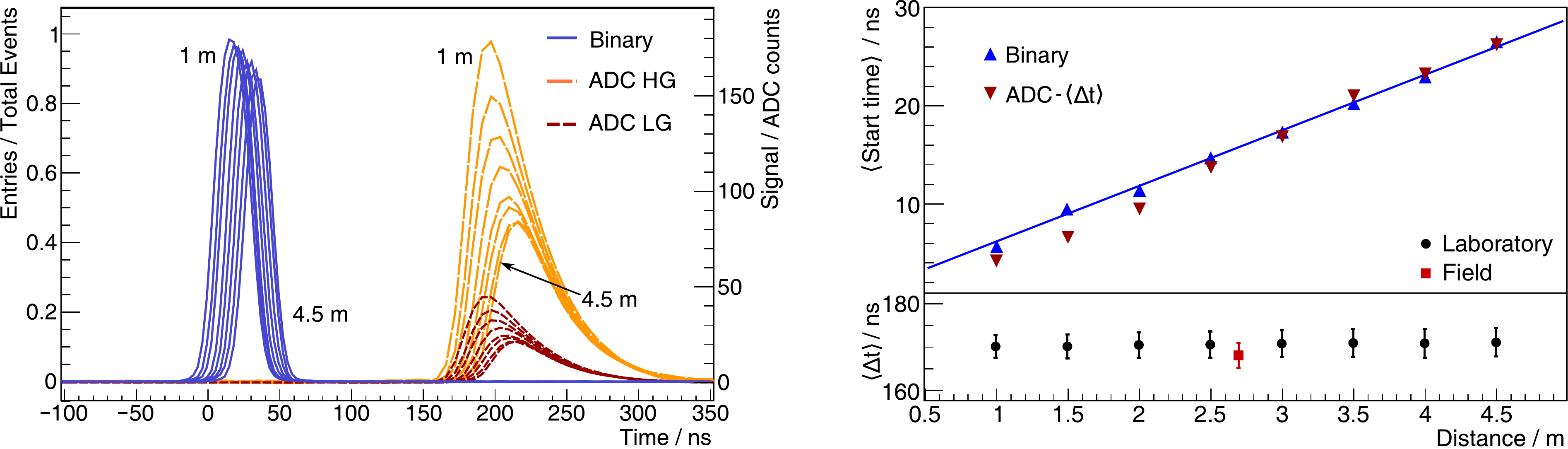}
	\end{subfigure}

	\caption{(Top) Schematics of the laboratory setup used to characterize the UMD traces. (Bottom-left) Average trace for the binary (solid-blue and left y-axis), ADC high- (long-dashed-orange and right y-axis), and low-gain (short-dashed-red) channels measured at different optical-fiber lengths (1 to 4.5\,m). (Bottom-right) Signal start time as a function of the optical-fiber length, for the binary (blue triangles) and ADC (red triangles) channels. The difference between the channel start times ($\Delta$t) measured in the laboratory (black circles) and the field (red square) is shown in the lower part.} 
	\label{fig:startTime}
\end{figure}

To study in detail how the two electronics relate one to another, we assembled in the laboratory a scintillator strip, identical to those used in the UMD, with an optical fiber coupled to a standard SiPM and read-out electronics, and a muon telescope movable at different positions along the scintillator strip. The muon telescope consists of two scintillator segments of 4\,cm$\times$4\,cm$\times$1\,cm, with a SiPM coupled to each segment. A schema of the setup is presented in the upper panel of Fig.~\ref{fig:startTime}. When a muon impinges on the telescope, the SiPM signals are correlated with an AND gate to inject a trigger signal in the UMD electronics. The muon also produces a signal in the scintillator strip: a sequence of positive samples in the binary mode and a waveform in the ADC mode is then obtained.

The signals read by the SiPM depend on the optical-fiber length between the SiPM and the position where the muon impinges on the scintillator. To characterize these signals, we measure 20000 muons produced at different positions on the scintillator strip, corresponding to lengths between 1\,m and 4.5\,m, in intervals of 0.5\,m. The average of these signals for each length is displayed in the bottom-left panel of Fig.~\ref{fig:startTime}. For the binary mode, each of the 2048-bit traces is summed and normalized to the total number of events (solid-blue curves and left y-axis) resulting in a histogram whose peaks correspond to the most probable bit to have a ``1''. For the ADC high- (long-dashed-orange curves) and low-gain (short-dashed-red curves) modes, the traces are averaged resulting in the mean waveform (right y-axis) of the muon signal. In both cases, we indicate the curves corresponding to the measurements at 1\,m and 4.5\,m. The general time shift between the ADC and the binary channels, introduced by different time response of the electronics, is apparent.

In the bottom-left panel of Fig.~\ref{fig:startTime}, two optical-fiber effects can be inferred. The light produced after a muon impinges on the scintillator strip is attenuated in the optical fiber and loses intensity as it propagates to the SiPM: the histogram peak of the binary channel and the maximum of the ADC waveform decrease as a function of the optical-fiber length. In addition, photons produced at different positions of the strip arrive at the SiPM at different times due to the delay introduced by the photon propagation in the optical fiber: the time of the histogram and waveform maximums is shifted when measuring at different positions on the strips. To better illustrate this fact, we present in the bottom-right panel of Fig.~\ref{fig:startTime} the mean start time of the binary channel as a function of the optical-fiber length (blue triangles). We define the start time of each individual trace as the time where the first ``1'' in the trace is found. The slope of this curve corresponds to the signal speed in the optical fiber, which we estimated, by fitting the data, to be $\mathrm{(5.67 \pm 0.05)\,ns/m}$, about 1\,/\,0.6\,c. This indicates that the photons in the optical fiber propagate at about 60\% of the speed of light in the vacuum\footnote{Due to the refractive index of the material (n = 1.60), the speed of light in the core of the optical fiber is 62.5\% of the speed in the vacuum. However, the speed of the signal is a convolution of the effect from the refractive index of the material and the path the photons follow when propagating due to total internal reflection in the fiber cladding. The selected fiber for the UMD is the Saint-Gobain BCF-99-29AMC.}.

Finally, we show in the bottom-right panel of Fig.~\ref{fig:startTime} the mean shift, $\langle\Delta$t$\rangle$, between the start times of the binary and ADC channels (black circles); the time shift was calculated event by event, and only averaged values are shown. The ADC start time is defined as the time at which the signal is five\,standard\,deviations above the baseline. We obtained that the time shift between channels is constant and does not depend on the position where the muon impinges on the scintillator strip. We also show the mean time shift calculated with data taken with a module deployed in the field (red square), as an example. In this case, as it is not possible to identify the position on the strip where the muons impinge on the scintillator, we assume the average optical-fiber length corresponding to the center of the module. The data from the field are well-matched with those taken in the laboratory. For completeness, the mean start times of the ADC channel, minus $\langle\Delta$t$\rangle$, are shown as red triangles in the panel above, to illustrate their consistency with the start times of the binary. As it will be explained in section \ref{sec:integrator}, characterizing the signal start time and the time shift between the channels constitutes a key point in the calibration of the ADC.

\section{SiPM calibration and setup of the binary acquisition mode}
\label{sec:SiPmcalibration}

Each SiPM of the UMD is an array of 1584 avalanche photo-diodes (also referred to as {\it cells}) operated in Geiger mode; the cells are distributed over an area of 2\,$\times$\,2\,mm$^2$. To set up the SiPM, a reverse bias voltage across the SiPM terminals ($\mathrm{V_{bias}}$) needs to be set. The minimum voltage for SiPMs to function corresponds to the breakdown voltage of the silicon junctions ($\mathrm{V_{br}}$). Then, the SiPM gain, noise and efficiency are proportional to the difference between $\mathrm{V_{bias}}$ and $\mathrm{V_{br}}$, commonly referred as {\it over-voltage} ($\mathrm{V_{ov}}$). When $\mathrm{V_{ov}}>0$, photons absorbed in the silicon produce electron-hole pairs, which trigger a diverging avalanche of carriers. Regardless of the number of photons impinging on a cell, this avalanche results always in the same macroscopic current, which only depends on the SiPM gain. Therefore, at the SiPM output, the signal amplitude and charge are proportional to $\mathrm{V_{ov}}$ and to the number of triggered cells, which we refer to as photo-equivalents (PE). Furthermore, since signal fluctuations are rather small, the amplitude (and charge) distributions for each number of PE are distinguishable one from another, making it possible to identify the number of PE in individual signals.

The noise in SiPMs is mainly caused by spurious pulses produced in absence of light. In silicon, there is a probability for carriers to be generated by thermal agitation; if an electron-hole pair is produced inside the active region of a cell, an avalanche may be initiated. We refer to the pulses that are not originated by incident radiation as dark counts. Furthermore, optical cross-talk between cells may occur when electrons are recombined during the avalanche emitting photons~\cite{sipms101}. These may escape from the triggered cell and hit neighbor cells producing simultaneous avalanches. As a result, an output of several PEs may be obtained, despite having only one cell triggered by incident radiation or thermal noise. In SiPMs, the dark-count rate and cross-talk probability are increasing function of the $\mathrm{V_{ov}}$ and of the temperature\footnote{The direct dependency of the cross-talk probability with the temperature is rather weak, if any, compared to most of the parameters in SiPMs.} ~\cite{sipms101,XTTemperature}.

\begin{figure*}[t]
\centering
\begin{subfigure}{0.95\textwidth}
\centering
\includegraphics[width=1.0\textwidth,clip]{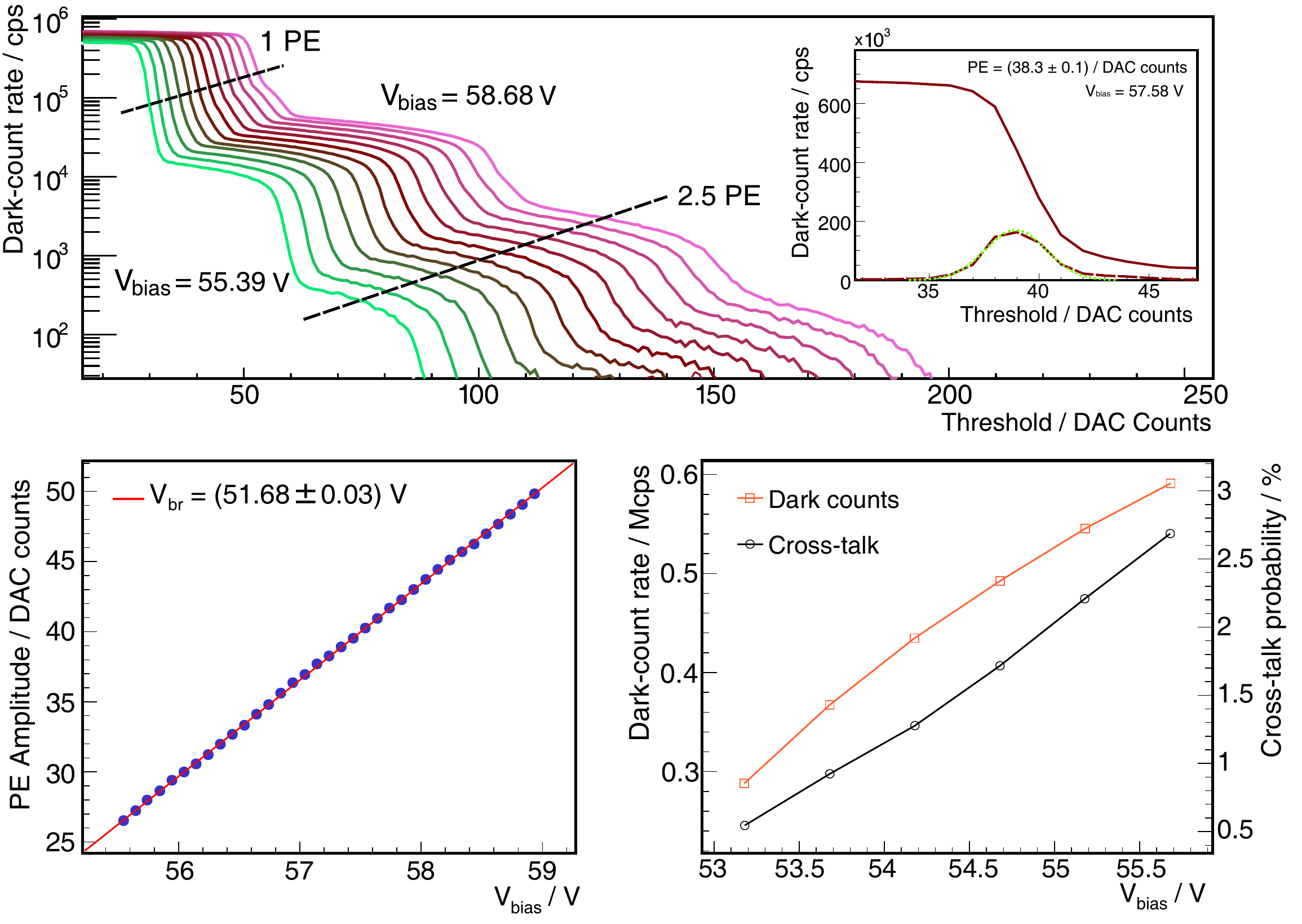}
\end{subfigure}
\caption{(Top) Dark-counts per second (cps) as a function of the discriminator threshold at different $\mathrm{V_{bias}}$ for an individual SiPM. The dark-count rates and PE amplitudes shift towards higher values when rising the $\mathrm{V_{bias}}$. In the inset, we show one curve at the middle of the displayed range in the main plot and its derivative fitted with a Gaussian distribution to obtain the corresponding PE amplitude. (Bottom-left) $\mathrm{V_{br}}$ determination for this specific SiPM. The linear extrapolation to 0 amplitude yields the value of $\mathrm{V_{br}}$. (Bottom-right) Dark-count rate (red squares) and cross-talk probability (black dots) as a function of the $\mathrm{V_{bias}}$.
\label{fig::calibration}}
\end{figure*}

The purpose of the SiPM calibration in the UMD is to equalize the gain (hence $\mathrm{V_{ov}=V_{bias}-V_{br}}$) of the 64 devices in each 10\,m$^2$ unit to achieve a uniform response in each scintillator module and, therefore, in the full UMD array. As $\mathrm{V_{br}}$ varies from one SiPM to another, the first step of the calibration is to determine the $\mathrm{V_{br}}$ of each. To this aim, we first measure the dark-count rates with different $\mathrm{V_{bias}}$. At each $\mathrm{V_{bias}}$, we vary the discriminator threshold of the binary channel, using a 10-bit digital-to-analog converter (DAC) integrated into the ASICs, and we count the number of SiPM signals per second above each threshold level. The measurements are performed at a reference temperature of 25\,$^\circ$C. As an example, we show in the top panel of Fig.~\ref{fig::calibration} the dark counts per second (cps) as a function of the discriminator threshold for one SiPM measured with different $\mathrm{V_{bias}}$, in which different plateaus can be clearly identified. The width of the plateau is related to the separation in amplitude between two consecutive integer numbers of PEs and corresponds to the transition between these signals.

Since the amplitude fluctuation of signals with a certain number of PEs is small, the rate of events varies slowly when the discriminator threshold is moved far from the mean signal amplitude. In contrast, when the discriminator threshold is near to the mean amplitude of the signal, the number of events changes rapidly and a transition between plateaus is observed. The first plateau corresponds to the transition between the baseline and 1\,PE signals, the second to the transition between 1 and 2\,PE signals, etc. The 1\,PE signal amplitude, which depends on $\mathrm{V_{bias}}$, can be easily identified in the middle between the first and second plateau. The sets of amplitude values for 1 and 2.5\,PEs are represented, as an example, in the figure with black dashed lines. To better illustrate the procedure for finding the PE amplitudes, we present in the inset a curve with its derivative fitted with a Gaussian distribution whose mean corresponds to the amplitude of 1\,PE for $\mathrm{V_{bias}}$ = 57.58\,V. By repeating this process with all the curves we can determine the signal amplitude of 1\,PE for different $\mathrm{V_{bias}}$.

The dark-count rate thus allows us to build the mean amplitude of the 1\,PE signal as a function of the $\mathrm{V_{bias}}$, 
as shown in the bottom-left panel of Fig.~\ref{fig::calibration}.
The mean amplitude increases linearly with $\mathrm{V_{bias}}$ as expected, and is zero for $\mathrm{V_{bias}} = \mathrm{V_{br}}$. With a linear fit to the measured amplitudes, we thus obtain $\mathrm{V_{br}}$ for each SiPM, which is $(51.68 \pm 0.04)$\,V in this example. The $\mathrm{V_{bias}}$ is then adjusted for each SiPM, so that all of them get the same $\mathrm{V_{ov}}$, making their response uniform.

Besides equalizing $\mathrm{V_{ov}}$ for the different SiPMs, the dark-count rate measurements allow us also to quantify the SiPM noise. We show in the bottom-right panel of Fig.~\ref{fig::calibration} the dark-count rate (red squares and y-axis on the left) and cross-talk probability (black dots and y-axis on the right) as a function of $\mathrm{V_{bias}}$. The dark-count rate is estimated as the rate of signals with amplitude $>0.5$~PE. In turn, the cross-talk probability is derived from the ratio between the rate of signals with amplitude >2\,PE and that with amplitude >1\,PE, due to the fact that events produced by thermal fluctuations can have more than 1\,PE amplitude only if cross-talk signals are produced effectively. As one can see from the figure, both the dark-count rate and the cross-talk probability depends on the $\mathrm{V_{bias}}$, given that, as mentioned before, the SiPM noise depends on the $\mathrm{V_{ov}}$.

Increasing the $\mathrm{V_{ov}}$ thus results, on the one hand, in a higher gain and efficiency, but also, on the other end, in higher noise. During the engineering array phase, the $\mathrm{V_{ov}}$ was optimized by balancing these parameters~\cite{AMIGASIPM}. With laboratory measurements we have shown that both a high efficiency and a sufficiently low noise are achieved with $\mathrm{V_{ov}} = 3.5$\,V ($\mathrm{V_{bias}} = 55.2$\,V in the figure). For the selected $\mathrm{V_{ov}}$, the typical value of cross-talk probability is about 2\% and that of dark-count rate, at 0.5\,PE, is between 450$\sim$850\,kcps, depending on the SiPM temperature. The discriminator threshold of the binary mode is then chosen to reduce the impact of this residual noise. Increasing the threshold results in a better background reduction but also loss of signal. 2.5 PE gives the best balance to reduce the dark-count rate measurements while keeping about 99\% of the muons from the muon telescope studies~\cite{UMDICRC}.

\begin{figure}[t]
    \centering
	\begin{subfigure}{1.00\textwidth}
		\centering
		\includegraphics[trim={0.0cm 0.0cm 0.0cm 0.0cm},clip,width=1.0\linewidth]{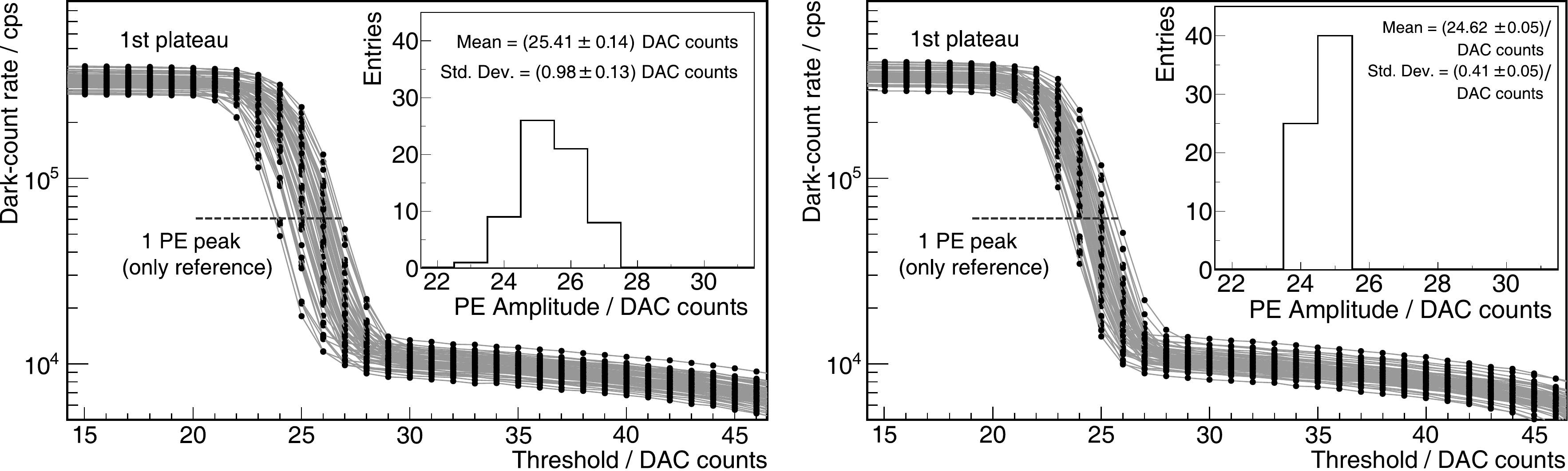}
	\end{subfigure}
    \caption{Dark-count rate as a function of the discriminator threshold for the 64 SiPMs of a typical scintillator module, before (left) and after (right) the calibration. Histograms of the signal amplitude corresponding to 1\,PE are also shown in the insets. Solid lines between data points are shown to ease the comparison.}
    \label{fig::curves}       
\end{figure}

The effect of the SiPM equalization can be appreciated in Fig.~\ref{fig::curves}, where we illustrate the result of the calibration performed on one scintillator module, as an example. We show the dark-count rate curves for the 64 SiPMs as a function of the discriminator threshold and, in the insets, the 1\,PE amplitude histograms, before (left) and after (right) calibration. The standard deviation of the histograms is reduced by a factor about two after calibration, which reflects the expected better uniformity of the detector response. Note that lines are displayed to ease the comparison between the curves, where points correspond to the data taken. After calibration, the dispersion of the gain is smaller than the minimum step of 1 DAC count, which corresponds to the DAC resolution. Note that while a large dispersion on the gain should be avoided, the chosen discriminator threshold in the middle of a plateau where the dark-count rate is rather constant minimizes the effect it has on rejecting dark counts. As can be inferred from the top panel of Fig.~\ref{fig::calibration}, variations of a few ADC counts in the gain have no impact in the performance and are acceptable for the full calibration procedure.

\begin{figure*}[t]
\begin{subfigure}{1.0\textwidth}
\centering
\includegraphics[width=1.00\textwidth,clip]{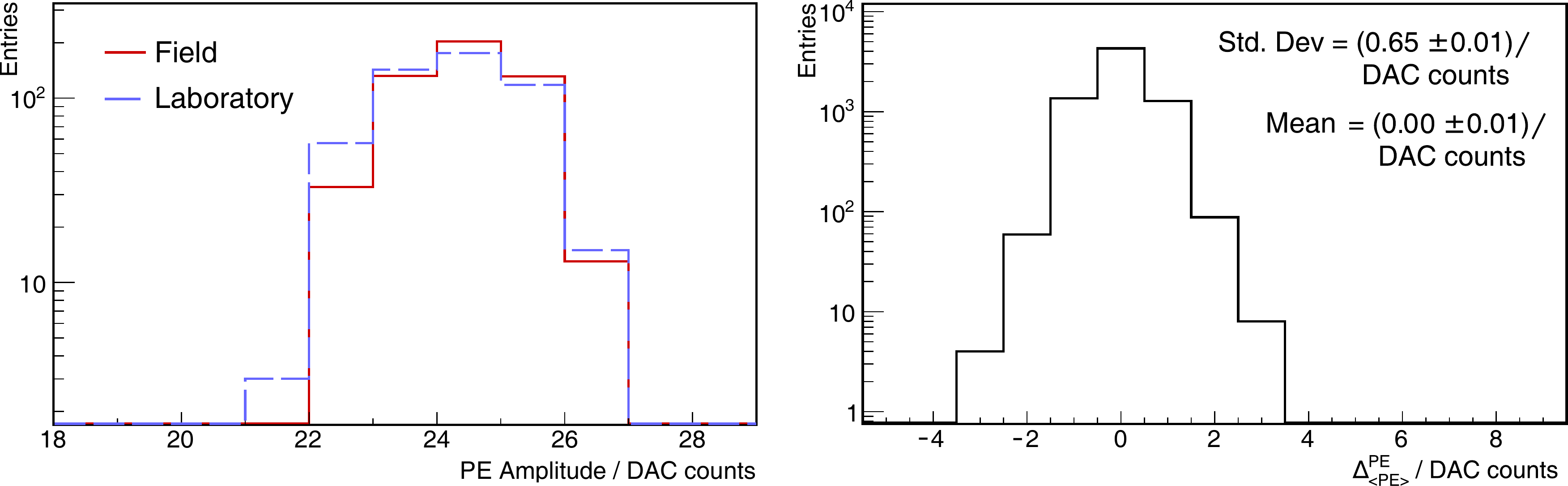}
\end{subfigure}
\caption{(Left) PE amplitude obtained in the laboratory and in the field for 512 SiPMs. (Right) PE amplitude for 6400 SiPMs; for each set of electronics, the mean amplitude of each module was centered to zero.
\label{fig::PEUni}}
\end{figure*}

The procedure described above was performed in the laboratory but it can be repeated at any time in the field, with the module removed from normal data acquisition. In fact, given that the strategy to calibrate the SiPM array and the binary channel is based on the analysis of thermal noise, performing the calibration in the laboratory or the field is, for any purpose, indistinct. This can be seen in the left panel of Fig~\ref{fig::PEUni} where we show the histograms of the PE amplitudes, obtained with dark-count measurements on a sample of 512 SiPMs, in the laboratory (dashed blue line) and in the field (solid-red line), after calibration. The two distributions are well-superimposed, the difference between the two measurements being always smaller than 0.5\,DAC\,counts. We compare the discriminator threshold levels using the mean of these histograms: in both cases the level of 2.5\,PE corresponds to 61\,DAC\,counts.

The calibration procedure has been applied, in the laboratory, to more than 6400 SiPMs: in the right panel of Fig.~\ref{fig::PEUni} we show that the distribution of the PE amplitudes minus the mean of each corresponding module results well-centered to zero. This shows that the calibration effectively allows us to obtain a uniform gain, the maximum deviation from the mean being of 3\,DAC\,counts (4.5 standard deviation), which, well within the acceptable range, does not have an impact on the detector performance. This calibration procedure guarantees a sufficient uniformity to work in all SiPMs of the UMD array.

\begin{figure}[t]
\vspace{-0.2cm}
\begin{subfigure}{1.0\textwidth}
\centering
\includegraphics[width=1.0\textwidth,clip]{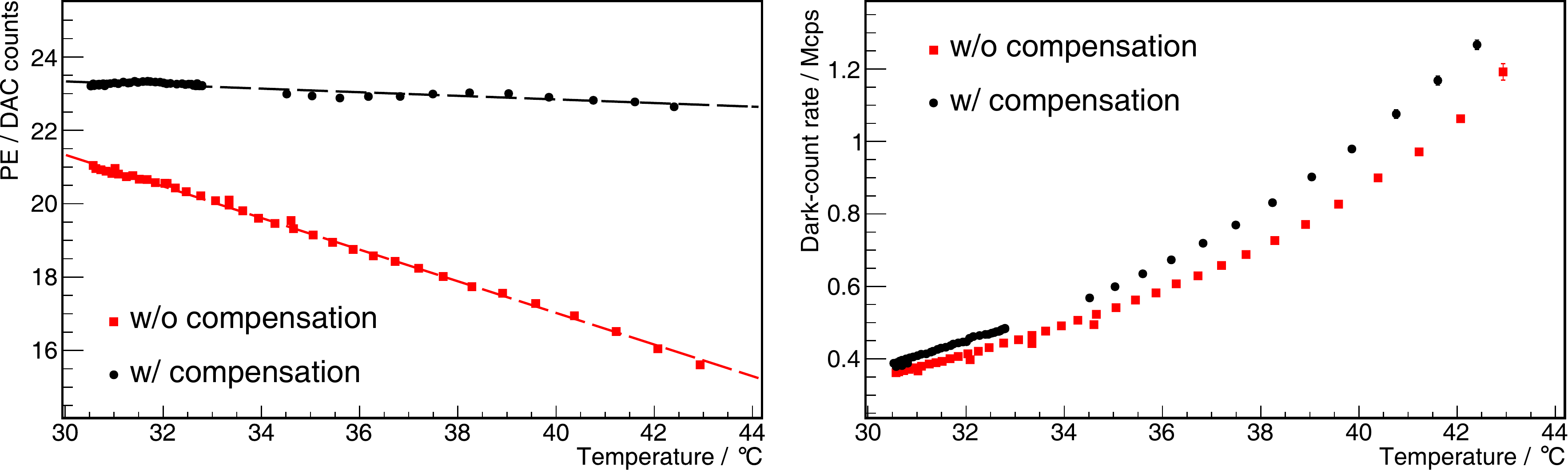}
\end{subfigure}
\caption{1\,PE amplitude (left) and dark-count rate (right) as a function of the temperature of the high-voltage source measured in the laboratory. We display the results for measurements with and without the compensation mechanism on. The fitted slopes for the PE amplitude data are (-0.049\,$\pm$\,0.003)\,PE\,/\,$^\circ$C and (-0.431\,$\pm$\,0.002)\,PE\,/\,$^\circ$C, respectively.
\label{fig::darkRate}}
\end{figure}

The SiPM gains, determined with the calibration, are however subjected to drifts, due to the well-known dependence of $\mathrm{V_{br}}$ with the temperature. The UMD operates in an outdoor ambient: even if it is buried 2.3\,m deep, it is exposed to temperature variations of about $20^\circ$ every year. The gain drifts must be compensated by readjusting $\mathrm{V_{bias}}$, to maintain the value of $\mathrm{V_{ov}}$ constant. Such an adjustment is done automatically, by exploiting a mechanism of temperature compensation that is integrated into the high-voltage source (C11204-01 by Hamamatsu). This mechanism adjusts the high voltage, i.e., $\mathrm{V_{bias}}$, according to the temperature variation: set at a reference value of $25\,^\circ \mathrm{C}$, it compensates by 54\,mV$/ ^\circ\mathrm{C}$, which is the temperature dependence of $\mathrm{V_{br}}$. In the left panel of Fig.~\ref{fig::darkRate}, we present the signal amplitude at 1\,PE (averaged over 64 SiPMs) as a function of the temperature of the high-voltage source measured in the laboratory without (red squares) and with (black circles) the temperature compensation. When the compensation mechanism is off and the temperature increases, then the $\mathrm{V_{br}}$ increases and, for a fixed $\mathrm{V_{bias}}$, the $\mathrm{V_{ov}}$ and the gain decrease. It is apparent how the PE amplitude is stabilized when the compensation mechanism is on. The remaining variation of about 2\,DAC\,counts in the whole temperature range is well within the acceptable range, and does not have an impact on the detector performance. 
In the right panel of Fig.~\ref{fig::darkRate} we show the average dark-count rate as a function of the temperature measured with and without the compensation mechanism. For comparison, the same temperature range is displayed for both measurements. Since the dark-count rate increases both with the SiPM temperature and $\mathrm{V_{ov}}$, when the compensation mechanism is off, two opposite effects contribute to the variation of the total rate: an increase with the temperature of the silicon and a decrease with the fall of the $\mathrm{V_{ov}}$. However, in the given ranges, the fluctuation of the dark-count rate due to the temperature variation (about 12\,$^\circ$C) is stronger than that produced by the variation of the $\mathrm{V_{ov}}$ (about 0.6\,V). Therefore, the dark-count rate fluctuation behaves similarly when measuring with and without the temperature compensation mechanism.

\begin{figure*}[t]
\begin{subfigure}{1.0\textwidth}
\centering
\includegraphics[width=1.0\textwidth,clip]{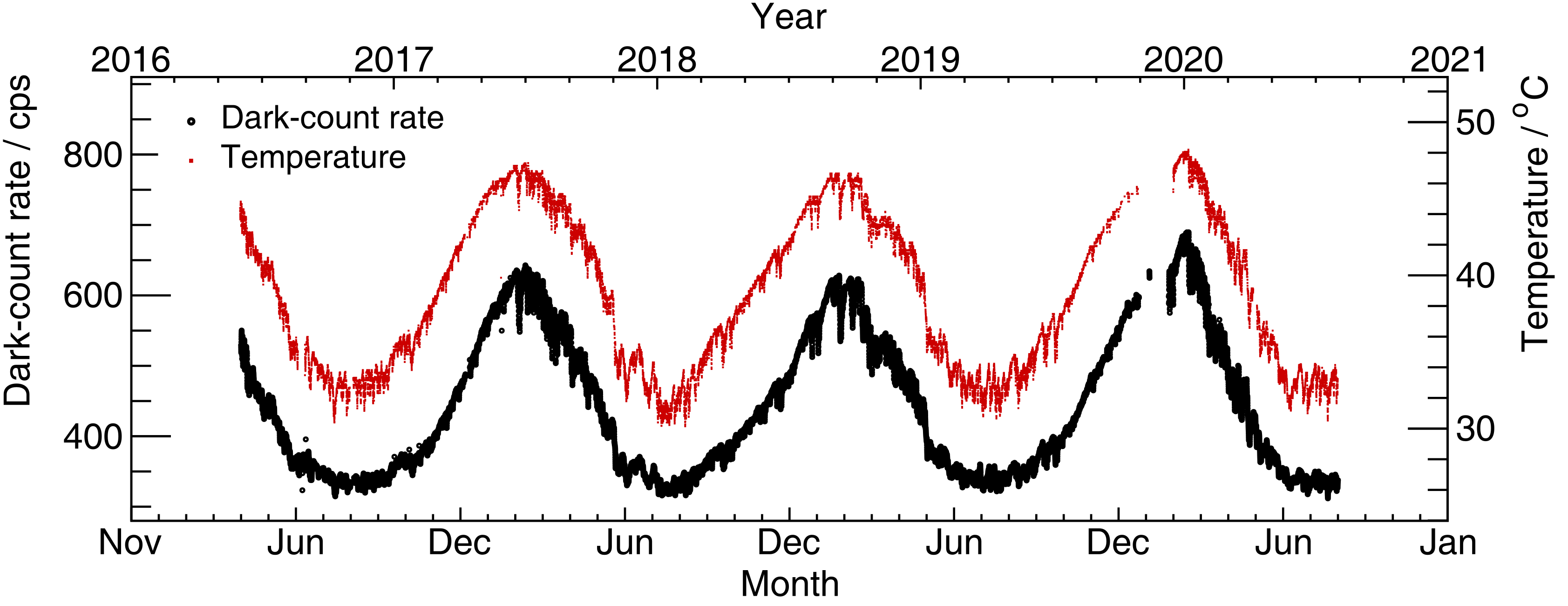}
\end{subfigure}
\caption{Mean rate per SiPM (left y-axis) and mean electronics temperature (right y-axis) averaged every hour in an example scintillation module in operation in the field. The gap between mid-December and January of 2020 corresponds to a period in which the corresponding position was not in acquisition.
\label{fig::darkRateField}}
\end{figure*}

The temperature range probed in the laboratory, between about $30^\circ$ and $42^\circ$, is contained within that experienced during operation in the field, as one can see in Fig.~\ref{fig::darkRateField}. 
The red squares show the variations of the temperature of the electronics (right y-axis), averaged every hour, in one scintillation module during more than three years of operation: the daily and yearly modulations are apparent. We also show the rate of signals with an amplitude $>2.5$\,PE averaged over all the SiPMs in the same module (black dots and left y-axis). Since the dark-count rate depends on the SiPM temperature, independently of the $\mathrm{V_{ov}}$, a modulation of the background rate is observed, despite the stability of $\mathrm{V_{ov}}$. The factor 2 in noise level between the warmest and coolest months of the years translates into a probability between 13\% and 25\% of having noise producing a ``1'' in the binary trace. This will not have a significant impact in the muon number estimation, since these signals can be rejected by setting a cut, explained in Section~\ref{subsec::muonSelection}.

\section{Calibration of the ADC acquisition mode}
\label{sec:integrator}

In the ADC channel, the number of muons is estimated by normalizing the charge of the recorded signal to the mean charge of a single muon. Therefore, the goal of the ADC calibration is to obtain the latter for each UMD module. To this aim, we exploit atmospheric muons, with the same approach used for the calibration of the WCD~\cite{WCDCalibration}. This approach is however more challenging for the UMD modules. 
As explained in Section~\ref{sec:electronics}, the trigger of each UMD module is not autonomous, but it depends on the first-level trigger of the associated WCD. The signals in the WCD and in the UMD are in fact mostly uncorrelated: in the majority of the events triggered by the WCDs, the signals in the UMD are largely due to noise. However, we show in this section how it is possible to select single-muon signals and use them to calibrate the ADC channel, thanks to the sufficiently high rate of the first-level WCD trigger (100\,triggers\,per\,second). The procedure aims at filling, during data acquisition, histograms of the muon-charge distribution, from which the average charge is extracted. Two ingredients need to be defined first: the strategy for the selection of muons, performed with the binary channel, and the calculation of their integrated charge, performed with the ADC channel.

\subsection{Muon selection criteria and charge calculation}
\label{subsec::muonSelection}

At the lowest trigger level, the bulk of the signals recorded in the ADC are due to SiPM dark counts. As no amplitude filter is implemented in the ADC acquisition mode (such as the discriminator threshold in the binary mode), each of the 64\,SiPMs contributes to the baseline fluctuation with more than 0.5\,Mcps of dark-count rate. To reduce such background while selecting signals due to muons, we make use of the binary channel. 
When a first-level trigger is received from the WCD, we build an internal second-level trigger based on the signal width, defined as the number of ``1''s in the binary trace. To define such a trigger, we first characterized the width of single-muon and noise signals by using a laboratory setup, including a muon telescope, similar to that described in section~\ref{sec:electronics} and shown in the upper panel of Figure~\ref{fig:startTime}. In this configuration, only one SiPM in the array is coupled to an optical fiber, with the other 63 being passive and contributing only with dark counts. The SiPM array and binary channel were calibrated as explained in Section~\ref{sec:SiPmcalibration}. 

\begin{figure}[t]
\centering
    \hspace{-0.2cm}
	\begin{subfigure}{0.495\textwidth}
		\centering
		\includegraphics[trim={0.0cm 0.0cm 0cm 0.0cm},clip,width=1.0\linewidth]{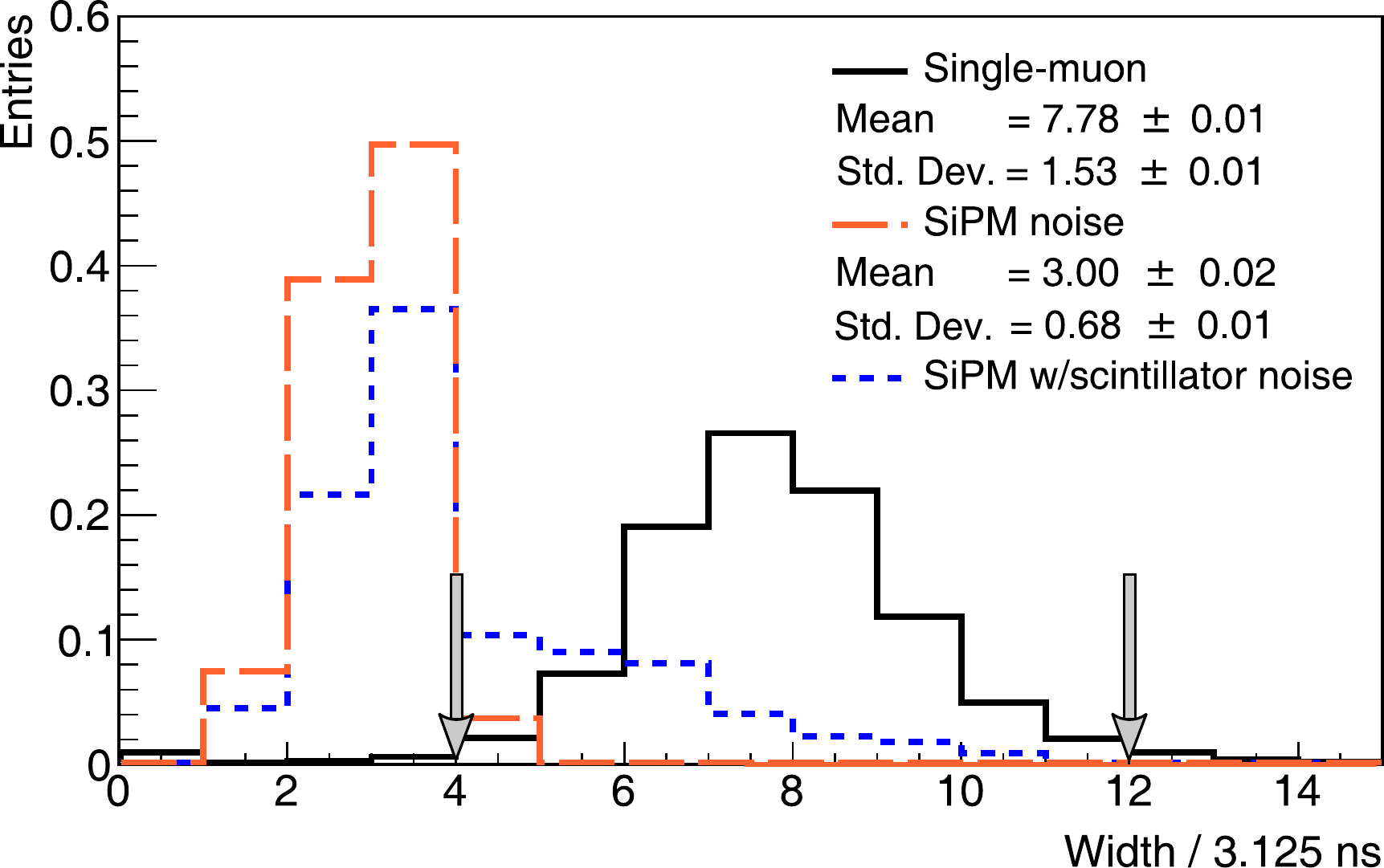}
	\end{subfigure}
	\hspace{0.1cm}
	\begin{subfigure}{0.49\textwidth}
		\centering
		\includegraphics[trim={0.0cm 0.0cm 0cm 0.0cm},clip,width=1.0\linewidth]{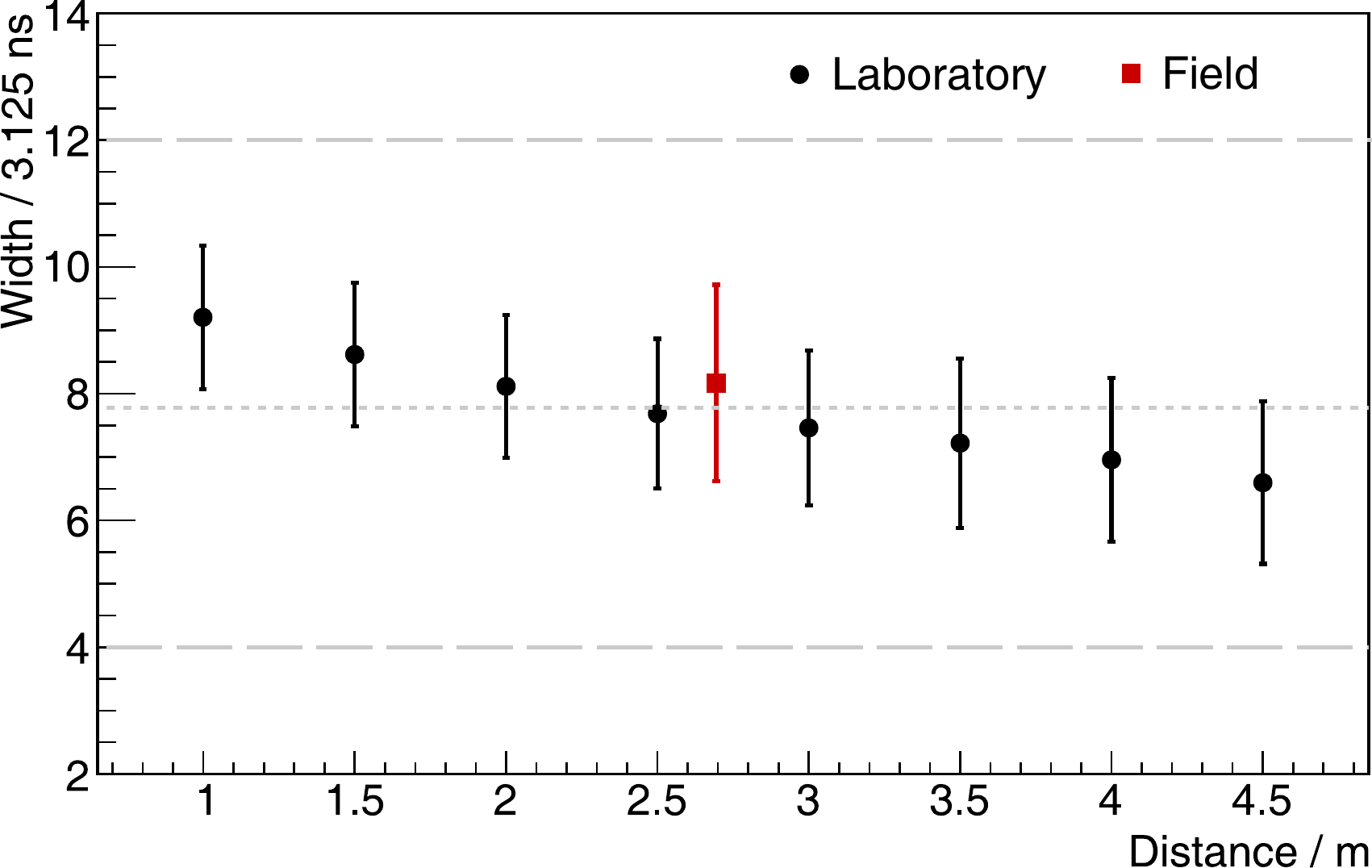}
	\end{subfigure}

\caption{Binary channel. (Left) histograms of signal widths due to single muons (black), SiPM noise (long-dashed red), and SiPM plus optical fiber-scintillator noise (short-dashed blue). (Right) mean width of single-muon signals as a function of the optical-fiber length. The one standard deviation from the mean is displayed as error bars. The red square corresponds to the result obtained with field data. The widths are indicated in number of samples (3.125\,ns).}
\label{fig::widthHistogram}       
\end{figure}

We used the muon telescope to collect 16000 signals at different positions of the scintillator strip, in the same way as explained in section~\ref{sec:electronics}. To characterize the muon signal, we defined a time window around the trigger time given by the telescope. Signals produced away from this window were classified as noise. The other 63 SiPMs were also turned on: when dark counts with amplitude $>2.5$~PE were produced, ``1''s in the corresponding binary traces were obtained. 
In the left panel of Fig.$~$\ref{fig::widthHistogram}, we display the histogram of the widths of the single-muon signal (solid-black line), based on the measurements made at all positions.
The spread of the distribution is in part an effect of the attenuation in the optical fiber, as illustrated in the right panel of Fig.~\ref{fig::widthHistogram}, where we show the behavior of the average width of the single-muon signal (black circles) as a function of the optical-fiber length. The error bars represent one standard deviation. One can clearly see that, as expected, longer (shorter) signals are detected closer to (farther from) the SiPM. 
In spite of such a spread, the distribution is well-separated from that due to noise signals, also shown in the left panel of Fig.$~$\ref{fig::widthHistogram}. The long-dashed red histogram corresponds to the distribution of the widths of the noise signals in the 63 SiPMs not coupled to the optical fiber, while the short-dashed blue one is due to noise-signals in the SiPM coupled to the optical fiber. The widths of the SiPM noise and of the muon signal are well-distinguished: the distribution relative to the muon signal has an average width of 7.8\,samples (24.4\,ns) with a standard deviation of 1.5\,samples (4.7\,ns) while that due to the SiPM-noise has an average width of 3.0\,samples (9.4\,ns) with standard deviation 0.7\,samples (2.2\,ns). Note that the noise in the SiPM coupled to the optical fiber (blue histogram) is the convolution of two contributions: the SiPM dark-count noise, the peak of which is at about 3\,samples, and a tail, with a mean at about 6\,samples, not related to the SiPMs, which we attribute to the optical-fiber/scintillator system~\cite{MINOS}. Even though the latter generates signals shorter than those due to muons, it cannot be efficiently rejected without losing muon signals. The effect of this irreducible background is however rather small, yielding a 5\% probability of over-counting a muon in a module when considering the whole trace of 6.4\,$\mu$s.

Based on these measurements, to select a muon we set a condition on the width of the signal, requiring more than four ``1''s (left arrow in the figure), which corresponds to about three\,standard\,deviation from the average duration. This cut rejects thus more than 95\% of the SiPM noise with less than 1\% of muon signal loss. Similarly, we also set an upper cut of 12\,samples (right arrow in the figure), which allows us to select about 99\% of the single-muon signals, avoiding biases in the muon charge estimation due to shower events with several particles. The two cuts are also shown as grey dashed lines in the right panel. To verify the applicability of these cuts to the UMD in the field, we measured the average width of binary signals in a module as an example. In the latter, we expect to observe wider signals than the mean found in the laboratory, since knock-on electrons from the soil also deposit energy in the scintillators, thus producing more light. Furthermore, the laboratory setup is more efficient at selecting vertical muons, which leave a smaller energy deposit than inclined muons as found in the field. The resulting measurement, shown in the figure as a red square, is still well-consistent with that obtained in the laboratory.

\begin{figure}[t]
\centering
    \hspace*{-0.2cm}
	\begin{subfigure}{.505\textwidth}
		\centering
		\includegraphics[trim={0.0cm 0.0cm 0.0cm 0.0cm},clip, width=1.00\linewidth]{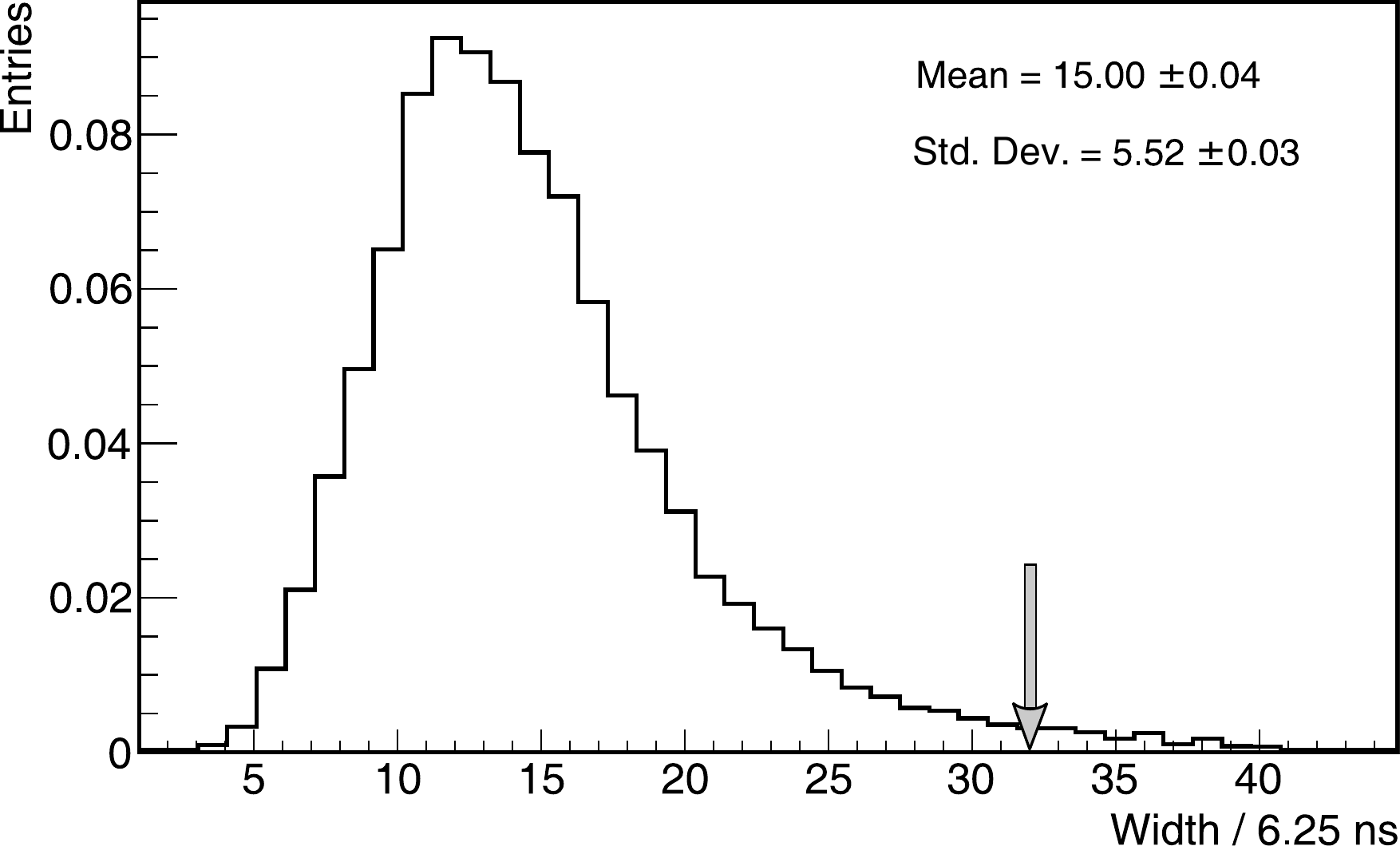}
	\end{subfigure}
	\hspace*{0.1cm}
	\vspace*{-0.07cm}
	\begin{subfigure}{.485\textwidth}
		\centering
		\includegraphics[trim={0.0cm 0.0cm 0.0cm 0.0cm},clip, width=1.00\linewidth]{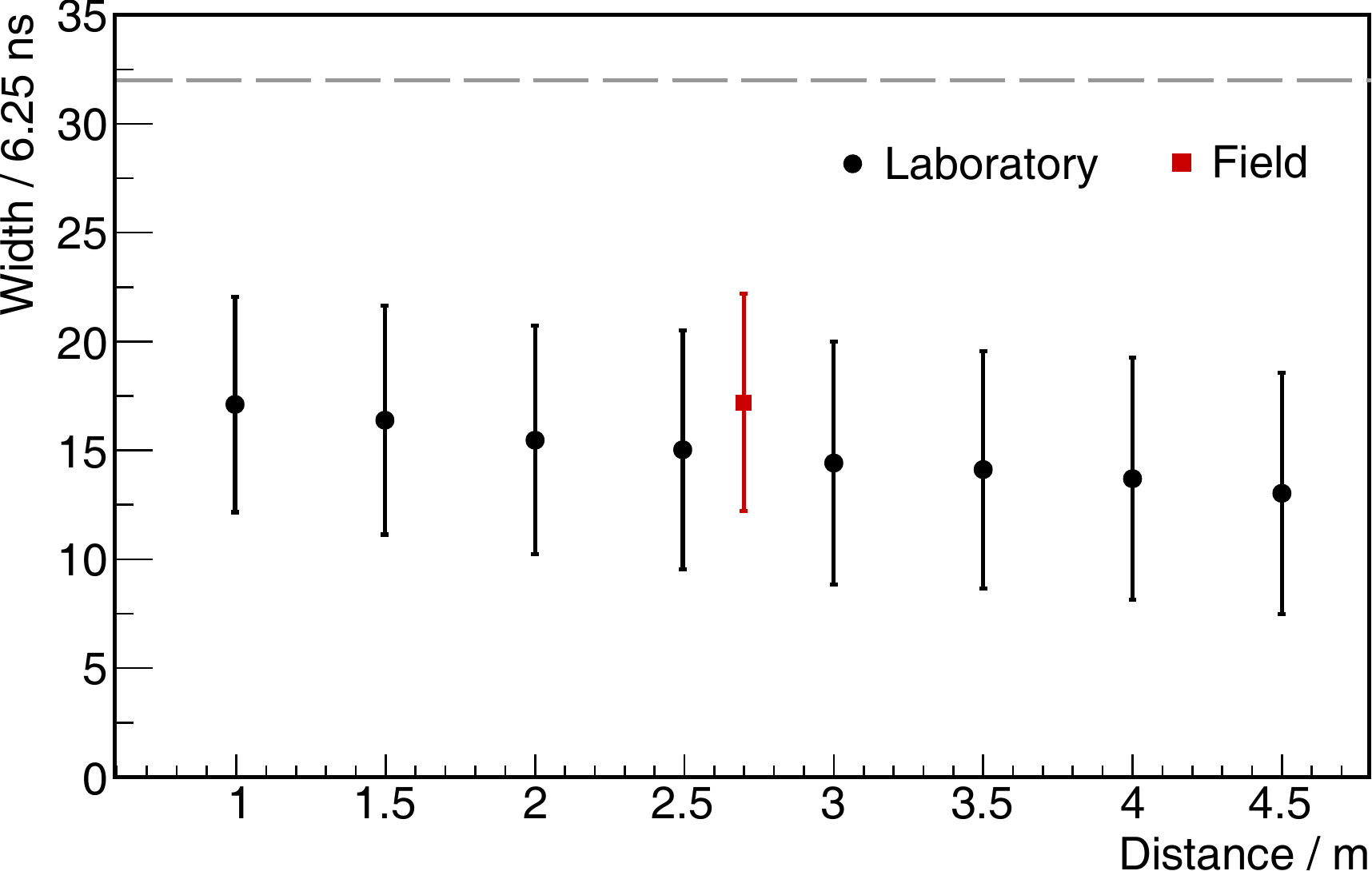}
	\end{subfigure}
\caption{ADC channel. (Left) histogram of the single-muon signal width measured with the ADC in the laboratory. (Right) Mean width of the single-muon signal as a function of the position of the impinging muon. Results from the laboratory (black circles) and field (red squares) with one standard deviation as error bars are presented. The widths are indicated in number of samples (6.25\,ns).}
\label{fig::chargeDistance}       
\end{figure}

The second ingredient required for the calibration procedure is the calculation of the integrated charge for the selected muon, by means of the ADC channel. To do that, an integration window needs to be optimized to reduce contributions from the noise due to the SiPM dark-count rate which causes fluctuations in the ADC baselines. 
By using the laboratory data acquired with the ADC channel, we studied the width of the muon signal, defined as the number of ADC bins with an amplitude above two times the standard deviation of the baseline. The distribution of the muon-signal widths is displayed in the left panel of Fig.~\ref{fig::chargeDistance}, with the right panel showing the behavior of the signal width (black circles) as a function of the optical-fiber length (error bars correspond to one standard deviation). Also, in this case, the signal becomes narrower as the distance increases, due to the light attenuation of the optical fiber. Based on this measurement, we defined the integration window for single-muon signals as three standard deviations from the mean width (grey arrow in the left panel, grey dashed line in the right one), which corresponds to 32\,samples (200\,ns). The red square corresponds to the mean width obtained from measurements in the field, showing its compatibility with the measurement in the laboratory.

\subsection{Filling of charge histograms}
\label{subsec::histoConstruction}

\begin{figure*}[t]
	\centering
	\vspace{0.5cm}
	\begin{subfigure}{1.0\textwidth}
		\centering
		\includegraphics[trim={0.0cm 0.0cm 0.0cm 0.0cm},clip, width=1.00\linewidth]{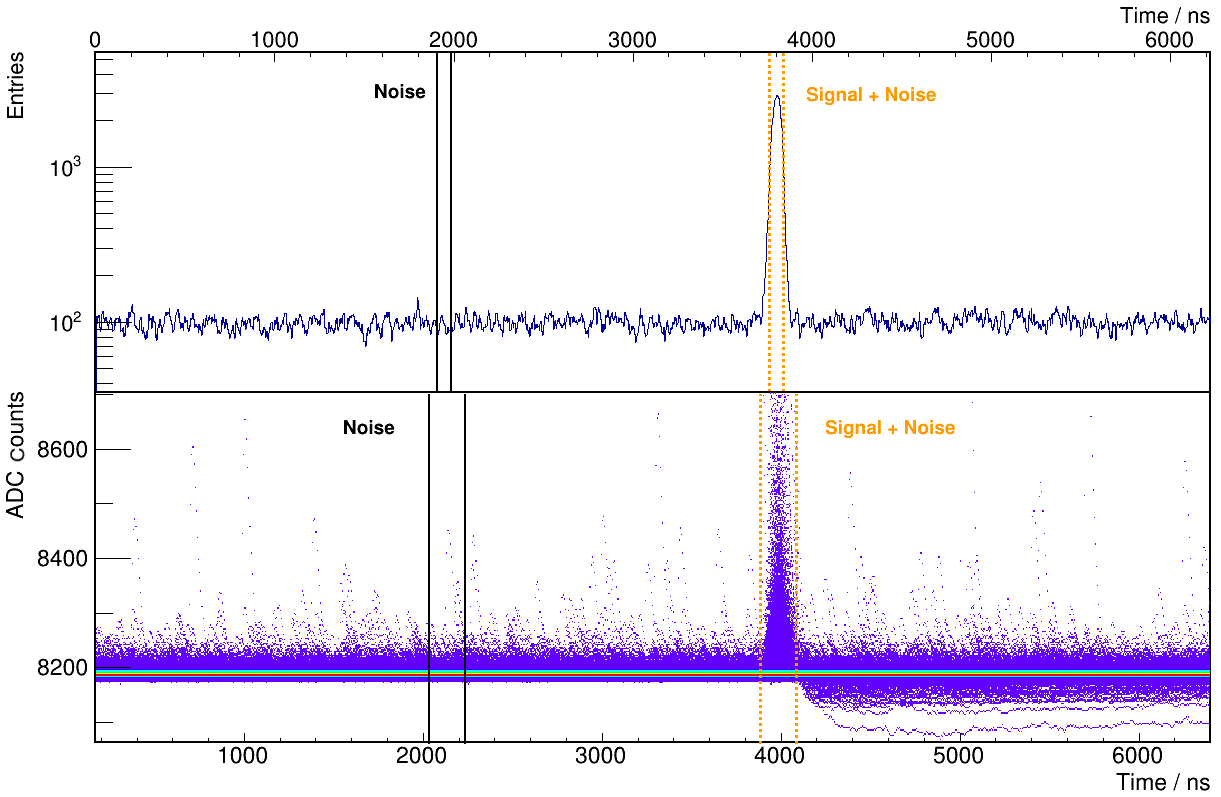}
	\end{subfigure}\\
	\caption{360000 background events which correspond to the equivalent of one hour of first-level triggered events. In orange (black) the time windows to search for muon (noise) signals are indicated. (Top) sum of binary channel traces. (Bottom) overlap of ADC channel traces. Note that the x-axis is shifted by about 170\,ns, the time shift between the channels presented in the bottom-right panel of Fig.~\ref{fig:startTime}.}
	\label{fig:windows}
\end{figure*}

The core of the calibration procedure is to periodically fill histograms of the signal charge produced by atmospheric muons. To explain the strategy, we start from Fig.~\ref{fig:windows}, where we show the traces from the binary channel (top) and the ADC (bottom), recorded following a WCD first-level trigger and summed over one hour of operation in the field (360000 events). The two histograms show similar features: a rather uniform distribution of signals along the trace, due to random detector noise, and an accumulation, due to particles, at a time that corresponds to that of the trigger from the WCD. Note that in the summed ADC traces undershoots are observed, after the trigger time, corresponding to high-amplitude signals, most likely due to air shower events.

To identify single-muon signals in the traces, we first search for ``1''s in a window around the trigger, illustrated with orange lines in the top panel. Then we apply the selection criterion described above, looking for a sequence of more than four and less than twelve ``1''s within this window. If this criterion is satisfied, we set as the start time of the binary trace the first ``1'' within the window, and then we locate the start-time of the ADC trace by including the information on the time shift between binary and ADC channels, presented in section~\ref{sec:electronics}. Finally, to extract the signal charge, we use the integration window of 32\,samples in the ADC trace (200\,ns), as explained above and illustrated by the orange lines in the bottom panel. As already mentioned, this procedure is effective to remove most of the noise from the SiPM, but not that from the optical fiber/scintillator system. Therefore, we repeat the above-described procedure, applying it in a time window separated from the trigger time window (represented by black lines in the bottom panel) where only noise contributes to the traces: we use this window to estimate the optical fiber/scintillator noise to subtract from the calibration data.

\begin{figure}[t]
	\centering
	
	%
	\vspace{1.5cm}
	\begin{subfigure}{1.05\textwidth}
		\centering
		\hspace{-0.5cm}
		\includegraphics[trim={0.0cm 0.0cm 0.0cm 0.0cm},clip, width=1.00\linewidth]{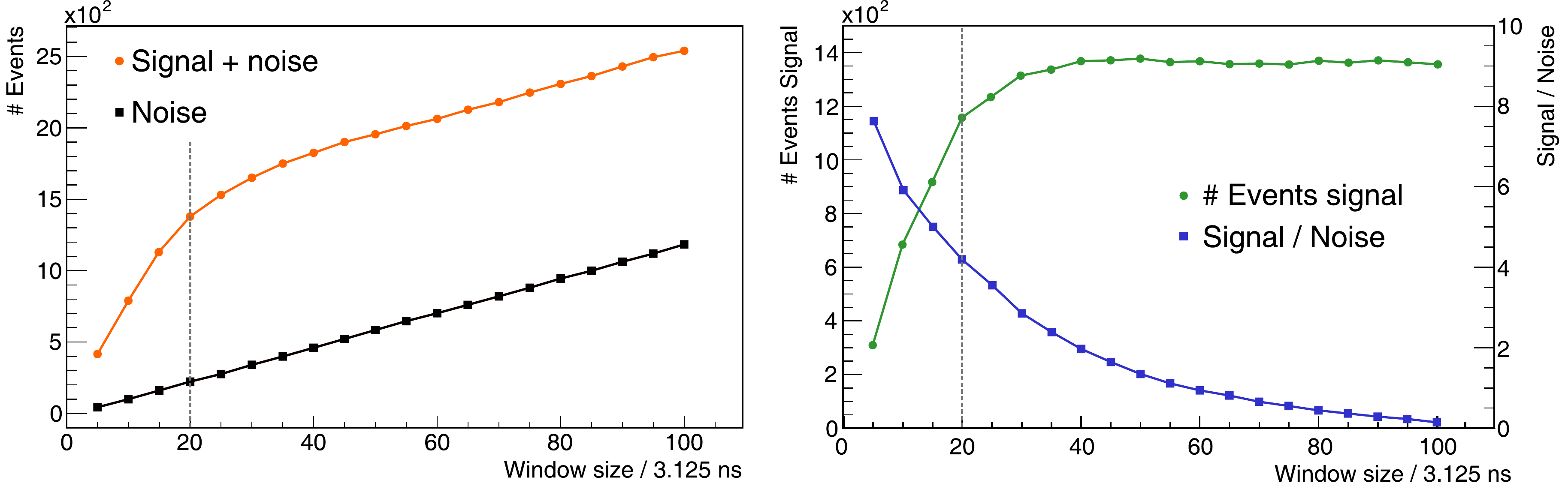}
	\end{subfigure} 
	\caption{Number of events in noise and signal-with-noise windows (left) and number of muon events (right) for each window size is displayed. The ratio between the number of signal and noise events is also presented.}
	\label{fig:nroEvents}
\end{figure}

The choice of the window size to search for muons and noise is important. Narrowing the size would reduce the number of noise signals, but it would also reduce the number of muon signals. To illustrate this effect, we show, in the left panel of Fig.~\ref{fig:nroEvents}, the number of events, satisfying the single-muon search criteria, in the noise window of the traces (black squares) and in the signal one (orange circles), as a function of the window size. As there are no muon signals in the noise window, the number of events grows linearly with its size. For the signal window, the number of events increases more rapidly than in the case of pure noise, up to about 35\,samples. Above this value, the curve becomes parallel to that due to only noise, thus indicating that no muons contribute any more to the number of events. This fact can be better observed in the right panel of Fig.~\ref{fig:nroEvents}, where we show the difference in the number of events between noise and signal windows (green circles): it increases for windows up to 35\,samples, then it becomes constant. In addition, we present the ratio between the number of signal and noise events (blue squares) as a function of the window size. The asymptotic behavior to zero indicates that, when increasing the window size, the number of noise events grows proportionally faster than the number of signal events. To implement the calibration procedure, we thus choose a window size of 20\,samples (indicated with gray-dashed lines), for which 80\% of the events are muons and 20\% are noise. Such a choice allows us to keep a large enough number of events, about 1200 for the analyzed period, minimizing the contamination due to noise.

\begin{figure}[t]
	\centering	
	\begin{subfigure}{0.6\textwidth}
		\centering
		\includegraphics[trim={0.0cm 0.0cm 0.0cm 0.0cm},clip, width=1.00\linewidth]{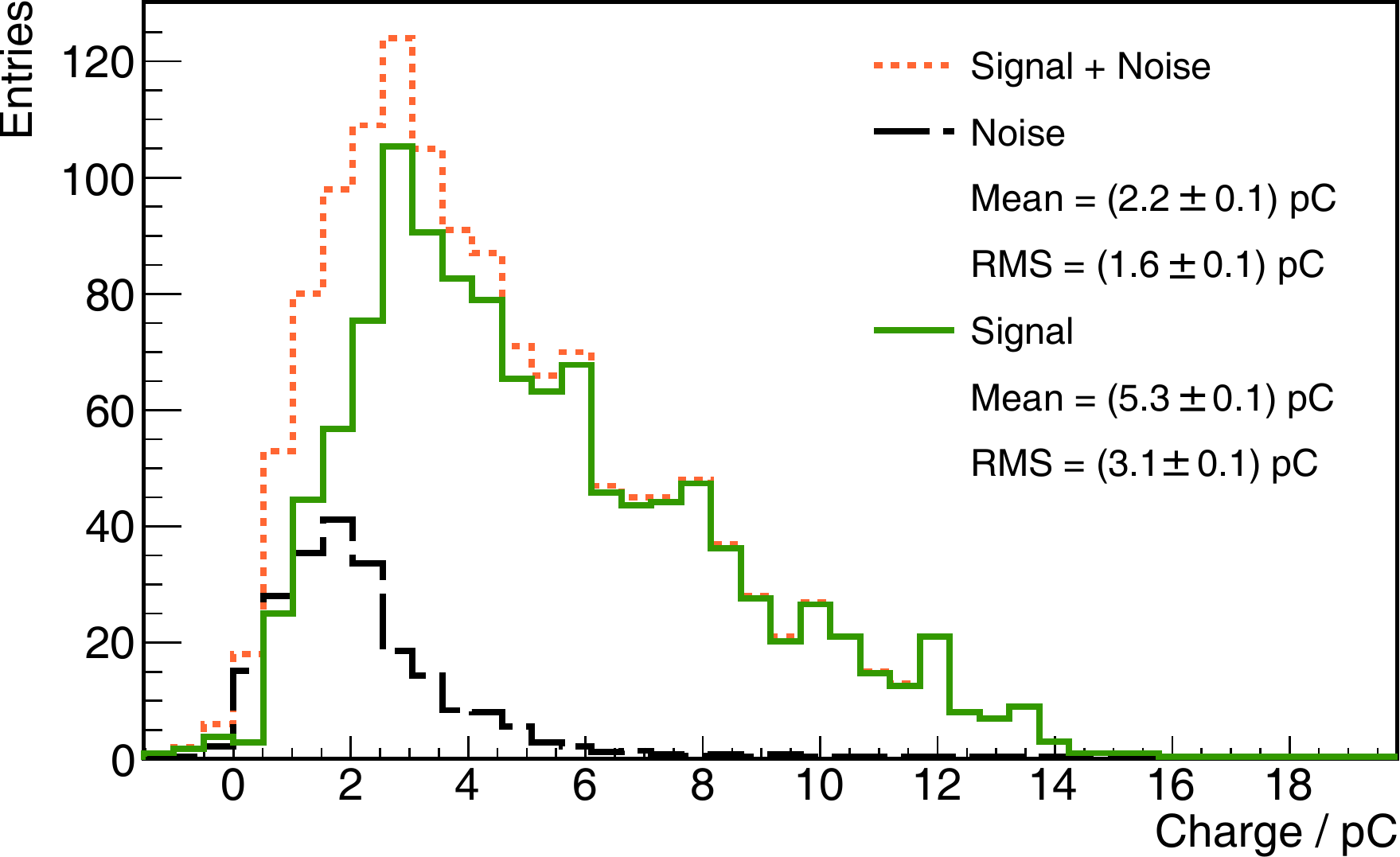}
	\end{subfigure}
	\caption{Charge histograms for a window size of 20\,samples. The calibration histogram (solid green) is filled by subtracting the charge histogram in the noise window from the charge histogram in the trigger window. 
	}
	\label{fig:charge}
\end{figure}

In Fig.~\ref{fig:charge} we present the muon charge histogram (green curve) obtained with the described procedure and a window of 20 samples. This histogram has been obtained by subtracting the noise histogram (dashed black) from the total charge histogram (dotted orange) that contains both signal and noise. The mean charge obtained in this case is (5.3 $\pm$ 0.1)\,pC with a standard deviation of (3.1 $\pm$ 0.1)\,pC. This corresponds to the mean charge deposited in the UMD by omnidirectional muons, and therefore includes the contribution of
secondary knock-on electrons produced in the 540 g/cm$^2$ overburden. The mean muon energy in showers measured with the UMD varies from about 2 GeV far from the shower core up to about 30 GeV when close. The contribution of knock-on electrons to the total charge is not expected to largely depend on the distance to the shower axis: on average about 10\% variation is expected in the estimation of the particle density for the most energetic muons close to the core with respect to those, less energetic, far from the impact point. For the analysis of cosmic-ray showers, dividing the total charge seen in a UMD module by the mean atmospheric muon charge measured in the same underground module will give, at first order, an estimation of the muon density at the surface.


To extract data for the ADC calibration, an algorithm is set up in each UMD module in the field, which runs in parallel to the air shower data acquisition. This algorithm is designed to select the traces that match the muon selection criteria, both for the signal and noise window. This information, along with the trigger timestamp, is stored in an external server, and the subsequent analysis is performed offline: a calibration histogram can be filled by selecting a time range of these data. To assess the performance of this calibration method under temperature variations, we have filled calibration histograms with data obtained with one module during several weeks in two periods with average electronic temperatures of 32$^\circ$C and 43$^\circ$C. In both cases, the mean of all histogram was (6.09 $\pm$ 0.01)\,pC with a standard deviation of (0.14 $\pm$ 0.01)\,pC and (0.12 $\pm$ 0.01)\,pC respectively, which translates into a fluctuation of the mean charge of about 2\%. Furthermore, the uniformity of the calibration has been tested over 21 modules in the field obtaining a mean charge of (6.09 $\pm$ 0.01)\,pC with a standard deviation of (0.5$\pm$ 0.1)\,pC, which indicates that the method is robust and its output quite uniform.

\section{Summary}
\label{sec:conclusion}
In this work, we have discussed the method developed to calibrate the underground muon detector of the Pierre Auger Observatory. A key characteristic of the scintillator-based detector is its high segmentation: each module is divided into 64 segments, coupled to a set of 64 SiPMs. The aim of the UMD is the direct measurement of the muon content in air showers with energy above $10^{16.5}$\,eV. This is achieved by implementing two acquisition modes: a binary mode, suitable for measuring low particle densities as found far from the shower core, and an ADC mode, tuned to measure high muon densities, such as those close to the shower core. For both channels, the ultimate scope of the calibration is the conversion from raw signals to the number of muons.

To develop the calibration method, correlations between the binary and ADC channels were first studied using a dedicated laboratory setup, consisting of a scintillator strip with optical fiber coupled to a SiPM and of a movable muon telescope. This allowed us, on the one hand, to study the response of the two channels to vertical muons at different positions, in terms of signal amplitude and start time. On the other hand, it allowed us to fully characterize the intrinsic time shift between the two channels.

Having characterized the relevant electronics features, we then proceed to explain the calibration of the binary channel. The first step is to equalize the SiPM gains and guarantee a uniform response of the UMD array which is of utmost importance since, in this channel, the number of muons is obtained by counting signals above a certain threshold, segment by segment. We implemented a method based on the measurement of dark-count rates, in which we first determine the reverse bias voltage for each SiPM, and then we set a threshold for the binary channel to 2.5\,PE, which filters most of the SiPM noise. By applying the method to more than 6400 SiPMs, we have verified that it allows us to achieve a PE amplitude with a maximum deviation from the mean of 3 DAC counts. Also, we have demonstrated that the performance of the calibration is the same when applied either in the laboratory or the field, thanks to the chosen strategy based on the analysis of thermal noise. Finally, we showed that the well-known drifts in the gains, due to temperature variations, are successfully stabilized by the temperature compensation mechanism integrated into the electronics, which reduces the temperature dependence by a factor of about 9.

Next, we presented the calibration of the ADC channel. In this case, the aim of the calibration is to provide the average charge of the signal of a single muon, so that the number of muons can be obtained by scaling the total charge measured in a UMD module. To this aim we use atmospheric muons, detected underground. 
Because the UMD is not triggered autonomously, but by the associated water-Cherenkov detector, only a part of the available signals correspond to muons and most of them correspond to uncorrelated noise. To overcome such a large background, we use the information in the binary channel to select muon signals. The selection criteria, based on the amplitude and the duration of the binary signals, have been defined after a study in the laboratory with a muon telescope. We have shown that they reject more than 95\% of the SiPM noise with about 1\% of muon signal loss. Once muons are selected with the binary channel, the start time of the signal in the ADC channel is identified and the charge is computed using an integration window which corresponds to the mean width of the single-muon signal plus three standard deviations, as measured in the laboratory and verified in the field. Finally, to obtain an unbiased estimation of the mean charge of single muons, we fill a charge histogram of the signal subtracted of that due to the detector noise. This method was tested in different UMD modules, also over time, denoting that the calibration is stable over time and uniform over the array.


In conclusion, we presented the methods to perform the UMD end-to-end calibration. We proved that these methods are robust to work with the high amount of channels to operate and stable under the temperature variations expected in the field. The methods described in this work will be applied to the 73 stations (219 modules) of the UMD array.


\section*{Acknowledgments}

\begin{sloppypar}
The successful installation, commissioning, and operation of the Pierre
Auger Observatory would not have been possible without the strong
commitment and effort from the technical and administrative staff in
Malarg\"ue. We are very grateful to the following agencies and
organizations for financial support:
\end{sloppypar}

\begin{sloppypar}
Argentina -- Comisi\'on Nacional de Energ\'\i{}a At\'omica; Agencia Nacional de
Promoci\'on Cient\'\i{}fica y Tecnol\'ogica (ANPCyT); Consejo Nacional de
Investigaciones Cient\'\i{}ficas y T\'ecnicas (CONICET); Gobierno de la
Provincia de Mendoza; Municipalidad de Malarg\"ue; NDM Holdings and Valle
Las Le\~nas; in gratitude for their continuing cooperation over land
access; Australia -- the Australian Research Council; Brazil -- Conselho
Nacional de Desenvolvimento Cient\'\i{}fico e Tecnol\'ogico (CNPq);
Financiadora de Estudos e Projetos (FINEP); Funda\c{c}\~ao de Amparo \`a
Pesquisa do Estado de Rio de Janeiro (FAPERJ); S\~ao Paulo Research
Foundation (FAPESP) Grants No.~2019/10151-2, No.~2010/07359-6 and
No.~1999/05404-3; Minist\'erio da Ci\^encia, Tecnologia, Inova\c{c}\~oes e
Comunica\c{c}\~oes (MCTIC); Czech Republic -- Grant No.~MSMT CR LTT18004,
LM2015038, LM2018102, CZ.02.1.01/0.0/0.0/16{\textunderscore}013/0001402,
CZ.02.1.01/0.0/0.0/18{\textunderscore}046/0016010 and
CZ.02.1.01/0.0/0.0/17{\textunderscore}049/0008422; France -- Centre de Calcul
IN2P3/CNRS; Centre National de la Recherche Scientifique (CNRS); Conseil
R\'egional Ile-de-France; D\'epartement Physique Nucl\'eaire et Corpusculaire
(PNC-IN2P3/CNRS); D\'epartement Sciences de l'Univers (SDU-INSU/CNRS);
Institut Lagrange de Paris (ILP) Grant No.~LABEX ANR-10-LABX-63 within
the Investissements d'Avenir Programme Grant No.~ANR-11-IDEX-0004-02;
Germany -- Bundesministerium f\"ur Bildung und Forschung (BMBF); Deutsche
Forschungsgemeinschaft (DFG); Finanzministerium Baden-W\"urttemberg;
Helmholtz Alliance for Astroparticle Physics (HAP);
Helmholtz-Gemeinschaft Deutscher Forschungszentren (HGF); Ministerium
f\"ur Innovation, Wissenschaft und Forschung des Landes
Nordrhein-Westfalen; Ministerium f\"ur Wissenschaft, Forschung und Kunst
des Landes Baden-W\"urttemberg; Italy -- Istituto Nazionale di Fisica
Nucleare (INFN); Istituto Nazionale di Astrofisica (INAF); Ministero
dell'Istruzione, dell'Universit\'a e della Ricerca (MIUR); CETEMPS Center
of Excellence; Ministero degli Affari Esteri (MAE); M\'exico -- Consejo
Nacional de Ciencia y Tecnolog\'\i{}a (CONACYT) No.~167733; Universidad
Nacional Aut\'onoma de M\'exico (UNAM); PAPIIT DGAPA-UNAM; The Netherlands
-- Ministry of Education, Culture and Science; Netherlands Organisation
for Scientific Research (NWO); Dutch national e-infrastructure with the
support of SURF Cooperative; Poland -Ministry of Science and Higher
Education, grant No.~DIR/WK/2018/11; National Science Centre, Grants
No.~2013/08/M/ST9/00322, No.~2016/23/B/ST9/01635 and No.~HARMONIA
5--2013/10/M/ST9/00062, UMO-2016/22/M/ST9/00198; Portugal -- Portuguese
national funds and FEDER funds within Programa Operacional Factores de
Competitividade through Funda\c{c}\~ao para a Ci\^encia e a Tecnologia
(COMPETE); Romania -- Romanian Ministry of Education and Research, the
Program Nucleu within MCI (PN19150201/16N/2019 and PN19060102) and
project PN-III-P1-1.2-PCCDI-2017-0839/19PCCDI/2018 within PNCDI III;
Slovenia -- Slovenian Research Agency, grants P1-0031, P1-0385, I0-0033,
N1-0111; Spain -- Ministerio de Econom\'\i{}a, Industria y Competitividad
(FPA2017-85114-P and PID2019-104676GB-C32, Xunta de Galicia (ED431C
2017/07), Junta de Andaluc\'\i{}a (SOMM17/6104/UGR, P18-FR-4314) Feder Funds,
RENATA Red Nacional Tem\'atica de Astropart\'\i{}culas (FPA2015-68783-REDT) and
Mar\'\i{}a de Maeztu Unit of Excellence (MDM-2016-0692); USA -- Department of
Energy, Contracts No.~DE-AC02-07CH11359, No.~DE-FR02-04ER41300,
No.~DE-FG02-99ER41107 and No.~DE-SC0011689; National Science Foundation,
Grant No.~0450696; The Grainger Foundation; Marie Curie-IRSES/EPLANET;
European Particle Physics Latin American Network; and UNESCO.
\end{sloppypar}

\bibliographystyle{JHEP}
\bibliography{UMD_strategies}{}

\begin{center}
\rule{0.1\columnwidth}{0.5pt}\,\raisebox{-0.5pt}{\rule{0.05\columnwidth}{1.5pt}}~\raisebox{-0.375ex}{\scriptsize$\bullet$}~\raisebox{-0.5pt}{\rule{0.05\columnwidth}{1.5pt}}\,\rule{0.1\columnwidth}{0.5pt}
\end{center}

\section*{The Pierre Auger Collaboration}

A.~Aab$^{80}$,
P.~Abreu$^{72}$,
M.~Aglietta$^{52,50}$,
J.M.~Albury$^{12}$,
I.~Allekotte$^{1}$,
A.~Almela$^{8,11}$,
J.~Alvarez-Mu\~niz$^{79}$,
R.~Alves Batista$^{80}$,
G.A.~Anastasi$^{61,50}$,
L.~Anchordoqui$^{87}$
% ...

{\footnotesize

\begin{description}[labelsep=0.2em,align=right,labelwidth=0.7em,labelindent=0em,leftmargin=2em,noitemsep]
\item[$^{1}$] Centro At\'omico Bariloche and Instituto Balseiro (CNEA-UNCuyo-CONICET), San Carlos de Bariloche, Argentina
\item[$^{2}$] Centro de Investigaciones en L\'aseres y Aplicaciones, CITEDEF and CONICET, Villa Martelli, Argentina
\item[$^{3}$] Departamento de F\'\i{}sica and Departamento de Ciencias de la Atm\'osfera y los Oc\'eanos, FCEyN, Universidad de Buenos Aires and CONICET, Buenos Aires, Argentina
\item[$^{4}$] IFLP, Universidad Nacional de La Plata and CONICET, La Plata, Argentina
\item[$^{5}$] Instituto de Astronom\'\i{}a y F\'\i{}sica del Espacio (IAFE, CONICET-UBA), Buenos Aires, Argentina
\item[$^{6}$] Instituto de F\'\i{}sica de Rosario (IFIR) -- CONICET/U.N.R.\ and Facultad de Ciencias Bioqu\'\i{}micas y Farmac\'euticas U.N.R., Rosario, Argentina
\item[$^{7}$] Instituto de Tecnolog\'\i{}as en Detecci\'on y Astropart\'\i{}culas (CNEA, CONICET, UNSAM), and Universidad Tecnol\'ogica Nacional -- Facultad Regional Mendoza (CONICET/CNEA), Mendoza, Argentina
\item[$^{8}$] Instituto de Tecnolog\'\i{}as en Detecci\'on y Astropart\'\i{}culas (CNEA, CONICET, UNSAM), Buenos Aires, Argentina
\item[$^{9}$] Observatorio Pierre Auger, Malarg\"ue, Argentina
\item[$^{10}$] Observatorio Pierre Auger and Comisi\'on Nacional de Energ\'\i{}a At\'omica, Malarg\"ue, Argentina
\item[$^{11}$] Universidad Tecnol\'ogica Nacional -- Facultad Regional Buenos Aires, Buenos Aires, Argentina
\item[$^{12}$] University of Adelaide, Adelaide, S.A., Australia
\item[$^{13}$] Universit\'e Libre de Bruxelles (ULB), Brussels, Belgium
\item[$^{14}$] Vrije Universiteit Brussels, Brussels, Belgium
\item[$^{15}$] Centro Brasileiro de Pesquisas Fisicas, Rio de Janeiro, RJ, Brazil
\item[$^{16}$] Centro Federal de Educa\c{c}\~ao Tecnol\'ogica Celso Suckow da Fonseca, Nova Friburgo, Brazil
\item[$^{17}$] Universidade de S\~ao Paulo, Escola de Engenharia de Lorena, Lorena, SP, Brazil
\item[$^{18}$] Universidade de S\~ao Paulo, Instituto de F\'\i{}sica de S\~ao Carlos, S\~ao Carlos, SP, Brazil
\item[$^{19}$] Universidade de S\~ao Paulo, Instituto de F\'\i{}sica, S\~ao Paulo, SP, Brazil
\item[$^{20}$] Universidade Estadual de Campinas, IFGW, Campinas, SP, Brazil
\item[$^{21}$] Universidade Estadual de Feira de Santana, Feira de Santana, Brazil
\item[$^{22}$] Universidade Federal do ABC, Santo Andr\'e, SP, Brazil
\item[$^{23}$] Universidade Federal do Paran\'a, Setor Palotina, Palotina, Brazil
\item[$^{24}$] Universidade Federal do Rio de Janeiro, Instituto de F\'\i{}sica, Rio de Janeiro, RJ, Brazil
\item[$^{25}$] Universidade Federal do Rio de Janeiro (UFRJ), Observat\'orio do Valongo, Rio de Janeiro, RJ, Brazil
\item[$^{26}$] Universidade Federal Fluminense, EEIMVR, Volta Redonda, RJ, Brazil
\item[$^{27}$] Universidad de Medell\'\i{}n, Medell\'\i{}n, Colombia
\item[$^{28}$] Universidad Industrial de Santander, Bucaramanga, Colombia
\item[$^{29}$] Charles University, Faculty of Mathematics and Physics, Institute of Particle and Nuclear Physics, Prague, Czech Republic
\item[$^{30}$] Institute of Physics of the Czech Academy of Sciences, Prague, Czech Republic
\item[$^{31}$] Palacky University, RCPTM, Olomouc, Czech Republic
\item[$^{32}$] CNRS/IN2P3, IJCLab, Universit\'e Paris-Saclay, Orsay, France
\item[$^{33}$] Laboratoire de Physique Nucl\'eaire et de Hautes Energies (LPNHE), Sorbonne Universit\'e, Universit\'e de Paris, CNRS-IN2P3, Paris, France
\item[$^{34}$] Univ.\ Grenoble Alpes, CNRS, Grenoble Institute of Engineering Univ.\ Grenoble Alpes, LPSC-IN2P3, 38000 Grenoble, France
\item[$^{35}$] Universit\'e Paris-Saclay, CNRS/IN2P3, IJCLab, Orsay, France
\item[$^{36}$] Bergische Universit\"at Wuppertal, Department of Physics, Wuppertal, Germany
\item[$^{37}$] Karlsruhe Institute of Technology, Institute for Experimental Particle Physics (ETP), Karlsruhe, Germany
\item[$^{38}$] Karlsruhe Institute of Technology, Institut f\"ur Prozessdatenverarbeitung und Elektronik, Karlsruhe, Germany
\item[$^{39}$] Karlsruhe Institute of Technology, Institute for Astroparticle Physics, Karlsruhe, Germany
\item[$^{40}$] RWTH Aachen University, III.\ Physikalisches Institut A, Aachen, Germany
\item[$^{41}$] Universit\"at Hamburg, II.\ Institut f\"ur Theoretische Physik, Hamburg, Germany
\item[$^{42}$] Universit\"at Siegen, Department Physik -- Experimentelle Teilchenphysik, Siegen, Germany
\item[$^{43}$] Gran Sasso Science Institute, L'Aquila, Italy
\item[$^{44}$] INFN Laboratori Nazionali del Gran Sasso, Assergi (L'Aquila), Italy
\item[$^{45}$] INFN, Sezione di Catania, Catania, Italy
\item[$^{46}$] INFN, Sezione di Lecce, Lecce, Italy
\item[$^{47}$] INFN, Sezione di Milano, Milano, Italy
\item[$^{48}$] INFN, Sezione di Napoli, Napoli, Italy
\item[$^{49}$] INFN, Sezione di Roma ``Tor Vergata'', Roma, Italy
\item[$^{50}$] INFN, Sezione di Torino, Torino, Italy
\item[$^{51}$] Istituto di Astrofisica Spaziale e Fisica Cosmica di Palermo (INAF), Palermo, Italy
\item[$^{52}$] Osservatorio Astrofisico di Torino (INAF), Torino, Italy
\item[$^{53}$] Politecnico di Milano, Dipartimento di Scienze e Tecnologie Aerospaziali , Milano, Italy
\item[$^{54}$] Universit\`a del Salento, Dipartimento di Matematica e Fisica ``E.\ De Giorgi'', Lecce, Italy
\item[$^{55}$] Universit\`a dell'Aquila, Dipartimento di Scienze Fisiche e Chimiche, L'Aquila, Italy
\item[$^{56}$] Universit\`a di Catania, Dipartimento di Fisica e Astronomia, Catania, Italy
\item[$^{57}$] Universit\`a di Milano, Dipartimento di Fisica, Milano, Italy
\item[$^{58}$] Universit\`a di Napoli ``Federico II'', Dipartimento di Fisica ``Ettore Pancini'', Napoli, Italy
\item[$^{59}$] Universit\`a di Palermo, Dipartimento di Fisica e Chimica ''E.\ Segr\`e'', Palermo, Italy
\item[$^{60}$] Universit\`a di Roma ``Tor Vergata'', Dipartimento di Fisica, Roma, Italy
\item[$^{61}$] Universit\`a Torino, Dipartimento di Fisica, Torino, Italy
\item[$^{62}$] Benem\'erita Universidad Aut\'onoma de Puebla, Puebla, M\'exico
\item[$^{63}$] Centro de Investigaci\'on y de Estudios Avanzados del IPN (CINVESTAV), M\'exico, D.F., M\'exico
\item[$^{64}$] Unidad Profesional Interdisciplinaria en Ingenier\'\i{}a y Tecnolog\'\i{}as Avanzadas del Instituto Polit\'ecnico Nacional (UPIITA-IPN), M\'exico, D.F., M\'exico
\item[$^{65}$] Universidad Aut\'onoma de Chiapas, Tuxtla Guti\'errez, Chiapas, M\'exico
\item[$^{66}$] Universidad Michoacana de San Nicol\'as de Hidalgo, Morelia, Michoac\'an, M\'exico
\item[$^{67}$] Universidad Nacional Aut\'onoma de M\'exico, M\'exico, D.F., M\'exico
\item[$^{68}$] Universidad Nacional de San Agustin de Arequipa, Facultad de Ciencias Naturales y Formales, Arequipa, Peru
\item[$^{69}$] Institute of Nuclear Physics PAN, Krakow, Poland
\item[$^{70}$] University of \L{}\'od\'z, Faculty of Astrophysics, \L{}\'od\'z, Poland
\item[$^{71}$] University of \L{}\'od\'z, Faculty of High-Energy Astrophysics,\L{}\'od\'z, Poland
\item[$^{72}$] Laborat\'orio de Instrumenta\c{c}\~ao e F\'\i{}sica Experimental de Part\'\i{}culas -- LIP and Instituto Superior T\'ecnico -- IST, Universidade de Lisboa -- UL, Lisboa, Portugal
\item[$^{73}$] ``Horia Hulubei'' National Institute for Physics and Nuclear Engineering, Bucharest-Magurele, Romania
\item[$^{74}$] Institute of Space Science, Bucharest-Magurele, Romania
\item[$^{75}$] University Politehnica of Bucharest, Bucharest, Romania
\item[$^{76}$] Center for Astrophysics and Cosmology (CAC), University of Nova Gorica, Nova Gorica, Slovenia
\item[$^{77}$] Experimental Particle Physics Department, J.\ Stefan Institute, Ljubljana, Slovenia
\item[$^{78}$] Universidad de Granada and C.A.F.P.E., Granada, Spain
\item[$^{79}$] Instituto Galego de F\'\i{}sica de Altas Enerx\'\i{}as (IGFAE), Universidade de Santiago de Compostela, Santiago de Compostela, Spain
\item[$^{80}$] IMAPP, Radboud University Nijmegen, Nijmegen, The Netherlands
\item[$^{81}$] KVI -- Center for Advanced Radiation Technology, University of Groningen, Groningen, The Netherlands
\item[$^{82}$] Nationaal Instituut voor Kernfysica en Hoge Energie Fysica (NIKHEF), Science Park, Amsterdam, The Netherlands
\item[$^{83}$] Stichting Astronomisch Onderzoek in Nederland (ASTRON), Dwingeloo, The Netherlands
\item[$^{84}$] Universiteit van Amsterdam, Faculty of Science, Amsterdam, The Netherlands
\item[$^{85}$] Case Western Reserve University, Cleveland, OH, USA
\item[$^{86}$] Colorado School of Mines, Golden, CO, USA
\item[$^{87}$] Department of Physics and Astronomy, Lehman College, City University of New York, Bronx, NY, USA
\item[$^{88}$] Louisiana State University, Baton Rouge, LA, USA
\item[$^{89}$] Michigan Technological University, Houghton, MI, USA
\item[$^{90}$] New York University, New York, NY, USA
\item[$^{91}$] Pennsylvania State University, University Park, PA, USA
\item[$^{92}$] University of Chicago, Enrico Fermi Institute, Chicago, IL, USA
\item[$^{93}$] University of Delaware, Department of Physics and Astronomy, Bartol Research Institute, Newark, DE, USA
\item[$^{94}$] University of Wisconsin-Madison, Department of Physics, Madison, WI, USA
\item[] -----
\item[$^{a}$] Fermi National Accelerator Laboratory, USA
\item[$^{b}$] Max-Planck-Institut f\"ur Radioastronomie, Bonn, Germany
\item[$^{c}$] School of Physics and Astronomy, University of Leeds, Leeds, United Kingdom
\item[$^{d}$] Fermi National Accelerator Laboratory, Fermilab, Batavia, IL, USA
\item[$^{e}$] also at Karlsruhe Institute of Technology, Karlsruhe, Germany
\item[$^{f}$] Colorado State University, Fort Collins, CO, USA
\item[$^{g}$] now at Hakubi Center for Advanced Research and Graduate School of Science, Kyoto University, Kyoto, Japan
\item[$^{h}$] also at University of Bucharest, Physics Department, Bucharest, Romania
\end{description}

}

\end{document}